\documentclass[twocolumn]{aastex631}

\usepackage{xcolor}

\newcommand\HI{H\,{\sc i}}

\shorttitle{Magnetic Sculpting of the Riegel--Crutcher Cloud}
\shortauthors{Franco et al.}

\graphicspath{{./}{figures/}}
\begin{document}

\title{When Magnetic Fields Sculpt the Sky: the Riegel–Crutcher cloud in optical polarization}

\correspondingauthor{Gabriel A. P. Franco}
\email{franco@fisica.ufmg.br}

\author[0000-0003-2020-2649]{Gabriel A. P. Franco}
\affiliation{Departamento de F\'\i sica -- ICEx -- UFMG, 
Av. Ant\^onio Carlos 6627, 30.270-901
Belo Horizonte, Brazil}

\author{Mayara Gomides}
\affiliation{Departamento de F\'\i sica -- ICEx -- UFMG, 
Av. Ant\^onio Carlos 6627, 30.270-901
Belo Horizonte, Brazil}

\author[0000-0002-7402-6487]{Zhi-Yun Li}
\affiliation{Astronomy Department, University of Virginia, Charlottesville, 
VA 22904, USA}
\affiliation{Virginia Institute of Theoretical Astronomy, University of Virginia, 
Charlottesville, VA 22904, USA}

\author[0000-0002-9650-3619]{F\'abio P. Santos}
\affiliation{Max-Planck-Institute for Astronomy, K\"onigstuhl 17, 
69117 Heidelberg, Germany}

\author{Farideh S. Tabatabaei}
\affiliation{Centre for Astrochemical Studies, 
Max-Planck-Institut f\"ur extraterrestrische Physik,
Gießenbachstraße 1, 85749 Garching bei M\"unchen, Germany
}

\begin{abstract} 
Filamentary structures are ubiquitous in the interstellar medium, yet the extent to which magnetic 
fields influence the morphology of cold atomic gas remains an open question. The nearby 
Riegel--Crutcher cloud, composed of long and narrow \HI\ filaments observed in self-absorption, 
provides a critical test case. We present the most extensive optical polarimetric survey of 
this region to date, comprising more than 90,000 high signal-to-noise stellar polarization 
measurements combined with \textit{Gaia} DR3 data. Using stellar polarization, extinction 
estimates, and archival Na\,{\sc i} absorption data, we locate the cloud at a distance of 
$150\pm15$ pc, consistent with that of the Pipe Nebula.

The plane-of-sky magnetic field traced by optical starlight polarization closely matches that 
inferred independently from \textit{Planck} 353\,GHz dust-emission polarization, revealing a 
coherent large-scale magnetic field across the region. A Rolling Hough Transform analysis shows 
that the \HI\ filaments are tightly aligned with this field orientation. Together, these results 
provide strong observational evidence that the structure of the cold neutral medium in the 
Riegel--Crutcher cloud is closely linked to a highly ordered magnetic field. 

This level of coherence supports a scenario in which magnetic fields play a dynamically important 
role in shaping the cloud structure, and suggests that the Riegel--Crutcher cloud is part of a 
larger magnetized complex influencing gas flows in the solar neighborhood.
\end{abstract}


\section{Introduction} \label{sec:intro}

Filaments are now recognized as a fundamental structural component of the interstellar 
medium (ISM), from diffuse gas to dense molecular clouds. Early studies already hinted at 
the filamentary nature of the ISM, but the advent of \textit{Herschel} confirmed their 
ubiquity and astrophysical importance, particularly in molecular clouds 
\citep{Menshchikov:2010, Molinari:2010, Schisano:2020}. More recently, 
high-resolution H\,{\sc i} surveys have revealed that even the diffuse atomic medium 
contains long, hair-like filamentary features \citep{Clark:2014, Kalberla:2016, 
Soler:2020, Soler:2022}. Many of these structures are oriented along the plane-of-sky 
(POS) magnetic field, suggesting that magnetic fields play a dominant role in shaping 
the cold neutral medium (CNM).

One of the most striking examples is the Riegel--Crutcher (R--C) cloud \citep{Riegel:1972}, 
mapped in H\,{\sc i} self-absorption (HISA) by \citet{McClure:2006}, who identified dozens of 
thin, coherent filaments. They compared filament orientation with the sparse optical 
polarimetry from \citet{Heiles:2000} and suggested magnetic regulation, but their analysis 
was limited by the small number of polarization measurements available at the time.

Here we present a new, deep optical polarimetric survey covering the main body of the 
R--C cloud. With more than 90,000 stars having high signal-to-noise polarization detections 
and accurate \textit{Gaia} distances, our data provide an unprecedented view of the magnetic 
field across this region. We used these data to (1) constrain the cloud distance, (2) map the 
POS magnetic field, (3) compare stellar and dust-emission polarization with \textit{Planck},
and (4) estimate the magnetic-field strength.

\section{Observational data}\label{sec:data}

\subsection{Data acquisition and reductions}\label{subsec:data_1}

Observations of optical polarization were collected in multiple observing campaigns 
conducted in 2015 and 2023. A polarimeter mounted onto the Cassegrain focus of the IAG 
Boller \& Chivens 60 cm telescope at the Observat\'orio do Pico dos Dias\footnote{Operated 
by Laborat\'orio Nacional de Astrof\'\i sica (LNA/MCTI -- Brazil)}, was used to get 
the data in the {\it V}-band of the Johnson-Cousins system. The polarimeter, a 
modified CCD camera, consists of a Savart prism and a rotating half-wave retarder that 
can be rotated in steps of $22\fdg5$.This arrangement provides two images of each object 
on the CCD with perpendicular polarizations \citep[see][for more details]{Magalhaes:1996}. 
A remarkable feature of this polarimeter is the simultaneous imaging of both the ordinary and 
the extraordinary beams, which allows for photon noise limited observations even under 
nonphotometric conditions, as well as cancellation of any sky polarization. Eight CCD images 
were taken for each field with the polarizer rotated through two modulation cycles of 
$0\degr$, $22\fdg5$, $45\degr$, and $67\fdg5$ in rotation angle. 

A total of 522 fields were observed, each covering an area of about $10' \times 10'$
covering a portion of the region of the R--C cloud studied by \citet{McClure:2006}. 
For each position of the half-wave plate, exposure times of 50 or 60 s were
used. In general, the longer exposure times were adopted in regions that were more strongly 
affected by interstellar absorption, in order to ensure adequate signal levels. The observing 
strategy was designed to sample a large number of independent regions of the sky; 
therefore, with the exception of a single field located close to the Galactic center position, 
the observed fields were arranged so that there was no overlap between them.

Data reduction was carried out using SOLVEPOL, a set of routines written by  
\citet{Ramirez:2017}  in Interactive Data Language (IDL).  After correcting each individual
image for zero level bias and flat-field, the routine identifies the corresponding pairs of 
stars and performs aperture photometry on them in each of the eight frames of a given field. 
The normalized linear polarization is calculated from a least-square solution by fitting 
the flux modulation of the two components which allows for the determination of the 
Stokes $Q$ and $U$ parameters, as well as the measured errors, $\sigma_P$, obtained from the
residuals of the observations at each wave plate position angle, $\psi_i$, with respect 
to the expected $\cos 4\psi_i$ curve. The polarization level and polarization angles,
measured from north to east, are calculated by

\begin{equation} \label{polarization_level}
P= \sqrt{Q^2 + U^2},
\end{equation}
and
\begin{equation} \label{polarization_angle}
\theta = \onehalf \arctan(U/Q).
\end{equation}
SOLVEPOL uses the {\it Astrometry.net} software \citep{Lang:2010} to calibrate the 
astrometry of the image to accurate celestial coordinates, which allows us to search the 
stars in the Guide Star Catalog \citep[GSC v2.3;][]{Lasker:2008} in order to perform the
magnitude calibration in the {\it V}-band \citep[see][for more details]{Ramirez:2017}.

\begin{figure}[b!]
\plotone{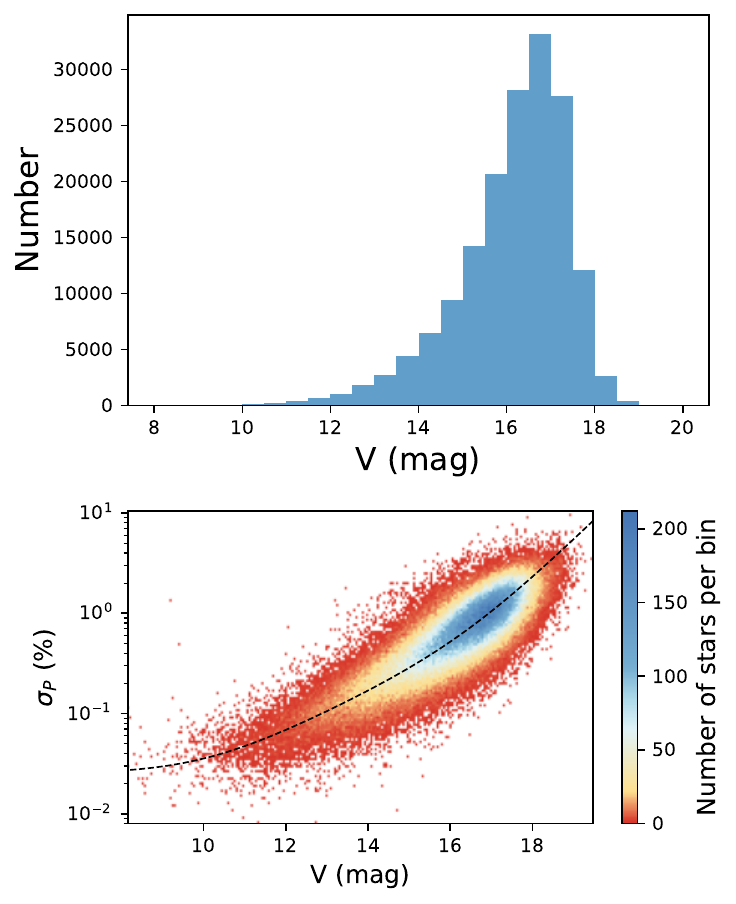}
\caption{\textit{Top:} distribution of V-band magnitudes for stars in our polarimetric sample 
that have counterparts in the \textit{Gaia} DR3 catalog.  
\textit{Bottom:} polarimetric uncertainty ($\sigma_P$) as a function of V-band magnitude. The color 
scale indicates the number of sources per bin in the density map. The distribution shows the expected 
behavior of the measurement uncertainties: a systematic floor of $\sim$0.03\% for bright stars, a 
photon-noise–dominated regime at intermediate magnitudes, and an increasing contribution from 
sky-background noise toward the faint end. The black dashed curve represents the empirical noise model
$\sigma_P^2 = \sigma_\mathrm{sys}^2 +\sigma_\mathrm{shot}^2 + \sigma_\mathrm{sky}^2$, where
$\sigma_\mathrm{shot} = b\times 10^{0.2(V-V_0)}$ and $\sigma_\mathrm{sky} = c \times 10^{0.4(V-V_0)}$,
with parameters $\sigma_\mathrm{sys}$ = 0.025\%, $b = 0.22$, $c = 0.10$, and $V_0$ = 14.7.}
\label{fig:hist_mag}
\end{figure}

Zero polarization standard stars were observed every run to check for any possible 
instrumental polarization, which proved to be small ($P \lesssim 0.1\%$), and in every
observing night 3 to 5 polarized standard stars were observed in order to determine
the reference direction of the polarizer.

\subsection{The obtained polarimetric data}\label{subsec:data_2}

With the aid of TOPCAT \citep{Taylor:2005}, we conducted a search in the \textit{Gaia DR3} 
catalog  \citep{Gaia:2023}, based on our stellar astrometry  (assuming a 1$\arcsec$ margin). 
Within this radius, the vast majority of the sources, i.e., about 94\% of the observed 
stars, have a single Gaia counterpart; in the cases where multiple Gaia sources are present, 
we adopted the likelihood-ratio method to estimate the probability that each candidate corresponds 
to the true counterpart  (see Appendix\,\ref{app:cross}).

The cross-match was successful for 166,706 stars, being that almost 
94,000 of them present $P/\sigma_P \ge 5$ corresponding to an uncertainty in polarization 
angle $\sigma_\theta \le 6\degr$ \citep{Naghizadeh:1993}. From the \textit{Gaia DR3} catalog we 
retrieved, when available, the trigonometric parallax and \textit{Gaia}'s photometry. In addition,  we
performed a second search in the \texttt{StarHorse}-based catalog \citep{Anders:2022} in 
order to retrieve distances and interstellar absorption. The cross-match for this second 
search was successful for 135,449 stars. Figure~\ref{fig:hist_mag} (top) shows the distribution 
of our {\it V}-band magnitudes and (botton) gives a density plot of the estimated 
polarization errors, $\sigma_P$,  as a function of the stellar magnitude. 
For bright sources, the uncertainties reach a nearly constant 
floor of $\sim$0.025\%, which is consistent with the level typically imposed by instrumental systematics 
and calibration residuals in well-behaved CCD polarimetry. At intermediate magnitudes, the 
uncertainties increase approximately as expected for photon 
shot noise, reflecting the decreasing signal-to-noise ratio of the stellar flux. Toward the faint end of the 
sample, the uncertainties rise more rapidly as the contribution from sky-background noise becomes 
increasingly important. This behavior was modeled using a simple empirical noise model that includes 
systematic, photon-noise, and sky-background components. The resulting curve reproduces well the ridge 
of the observed distribution.

\begin{figure}[htb!]
\plotone{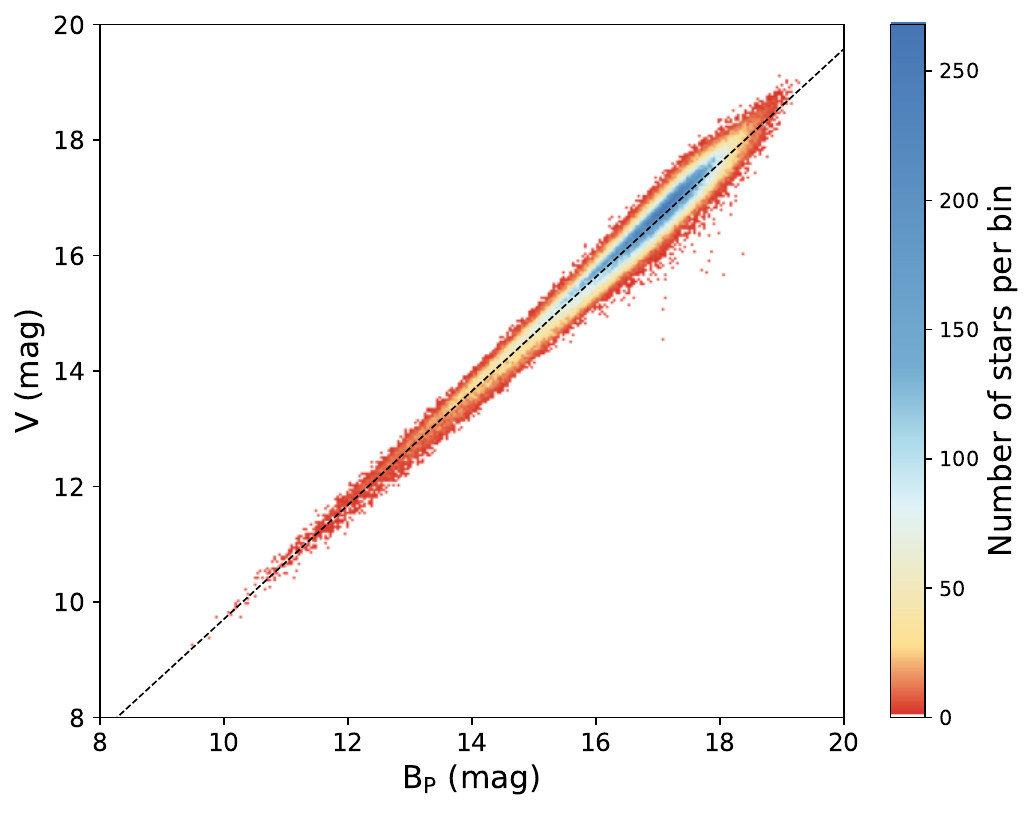}
\caption{Comparison between the $V$-band magnitudes estimated from our polarimetric observations 
and the \textit{Gaia} DR3 B$_{\rm P}$-band photometry for the cross-matched stellar sample. The 
color scale indicates the density of points, and the dashed line shows the best-fit linear regression, 
$V = 0.99\,\mathrm{B_P} - 0.17$. The tight correlation and small scatter demonstrate the reliability 
of the $V$-band magnitude calibration adopted for our optical polarization catalog.
\label{fig:bp_V}}
\end{figure}

In Fig.~\ref{fig:bp_V} we present a comparison between our estimated $V$-band 
photometry and the retrieved $B_{\rm P}$-band from \textit{Gaia}. The excellent agreement 
between these two photometries attests in favor of the quality of our observational data 
and gives us confidence to the cross-match with the \textit{Gaia} catalog.
 A small number of outliers are present, typically showing Gaia magnitudes fainter than the 
estimated V magnitudes. Inspection of these sources indicates that they generally correspond to 
stars with low signal-to-noise polarization measurements ($P/\sigma_P < 5$), which are not included in the 
main analysis presented in this work. Furthermore, these objects typically exhibit very small Gaia parallaxes, 
placing them at large distances and outside the distance range considered in this study. Consequently, 
possible mismatches among these faint sources do not affect the final sample used in the scientific analysis.

\section{Distance to the Riegel-Crutcher cloud}\label{sec:dist}

\begin{figure}
\plotone{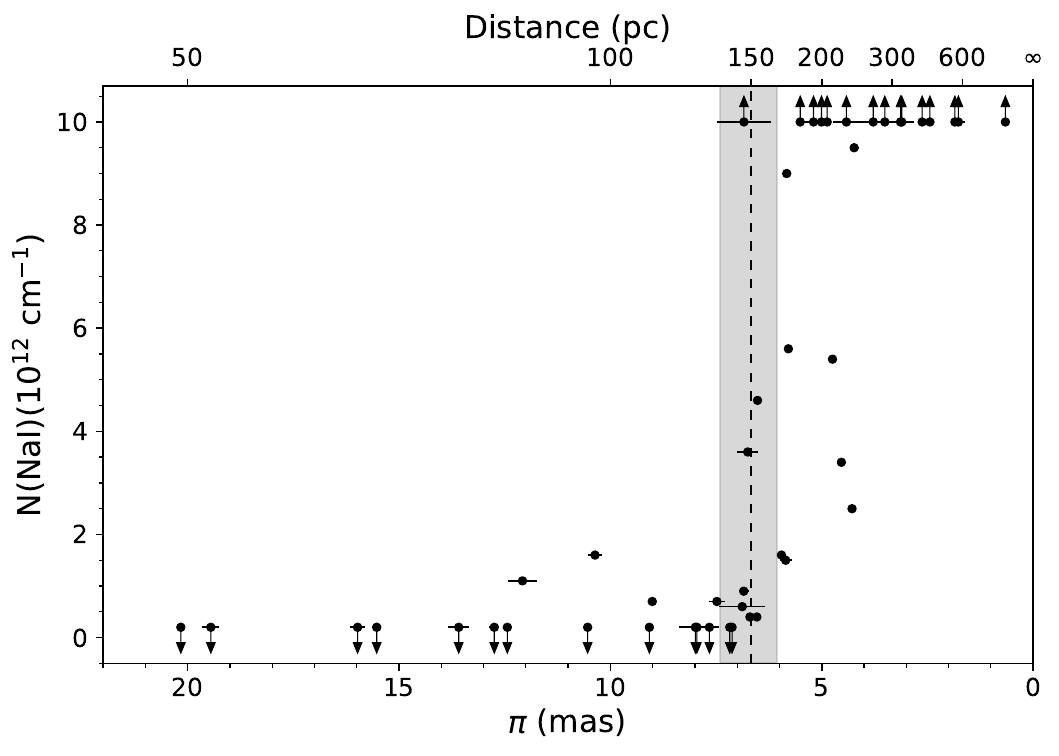}
\caption{Na\,\textsc{i} D-line column density from \citet{Crutcher:1984} as a function of stellar distance 
estimated using \textit{Gaia} DR3 parallaxes. Horizontal bars represent the distance uncertainties for 
each star, while vertical arrows indicate lower (upward) or upper (downward) limits to the 
Na\,\textsc{i} column density. The vertical gray band marks the inferred distance to the R--C cloud 
($150 \pm 15$ pc). The systematic rise in Na\,\textsc{i} absorption beyond $\sim$140--160 pc indicates 
the onset of the cold foreground material associated with the R--C cloud, confirming that it lies 
farther than earlier estimates based on spectrophotometric distances.
\label{fig:naixdist}}
\end{figure}

A distance of 125$\pm$25 pc to the R--C cloud was determined by \citet{Crutcher:1984} using observations 
of the interstellar Na\,{\sc i} D lines toward 49 early type stars in the direction of the cloud. The distance to these 
stars was estimated taking into account their stellar spectral type, visual magnitude, and color excess. We
used \textit{Gaia}'s parallax ($\pi$) in order to reanalyze the distribution of the Na\,{\sc i} D column density as a 
function of the stellar distance. The new diagram is presented in Fig.\,\ref{fig:naixdist}, where the vertical
dashed line was placed at a parallax $\pi \approx 0.667$ mas, corresponding to a distance of 150 pc, the
gray strip denotes the interval 150$\pm$15 pc. Based on this diagram we may suppose that the R--C cloud is 
likely to be a little farther away than the distance suggested by \citet{Crutcher:1984}.

\begin{figure}
\plotone{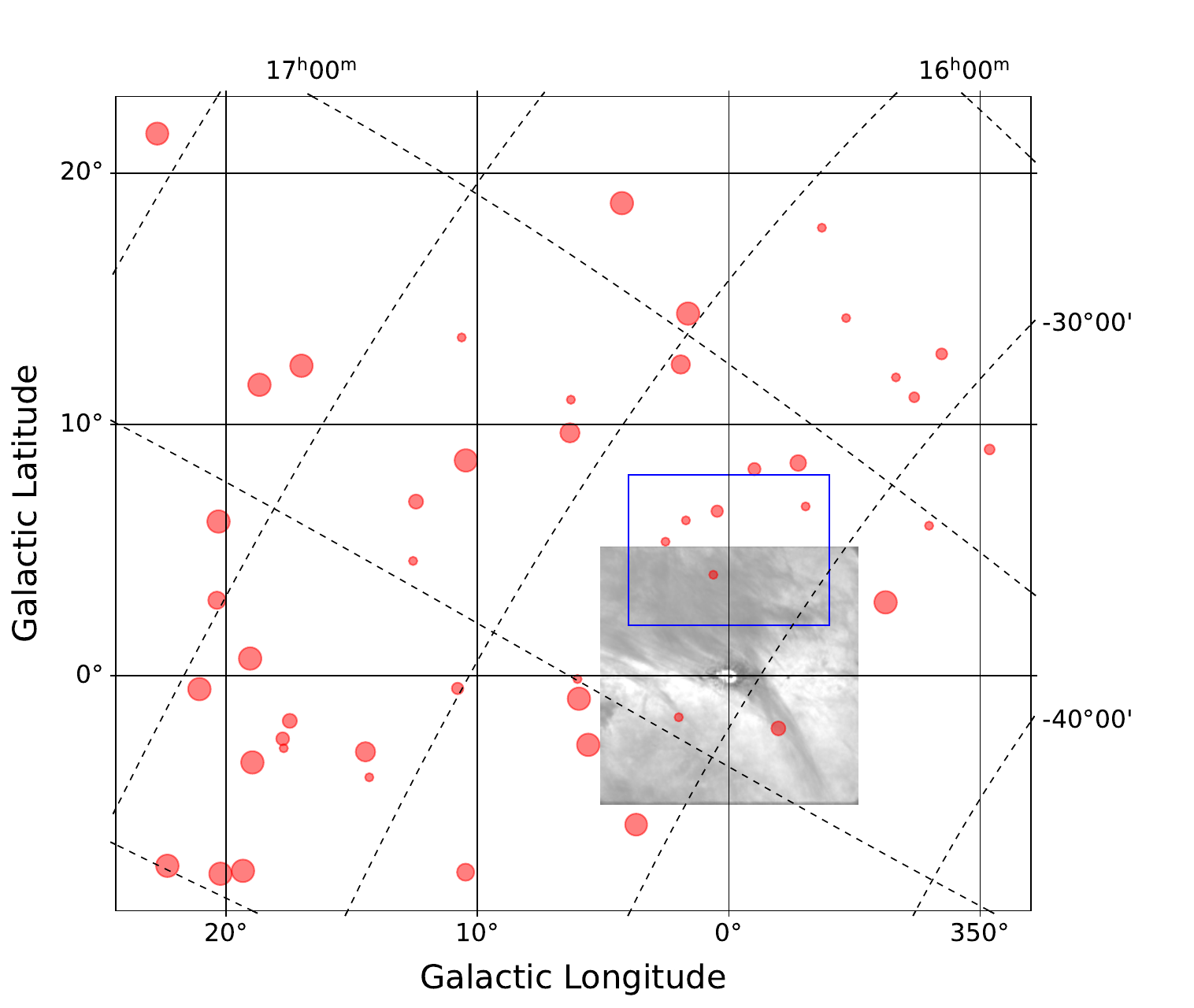}
\caption{Spatial distribution of the stars observed by \citet{Crutcher:1984} in the direction of the 
R--C cloud. The size of each circle is proportional to the Na\,\textsc{i} column density measured 
toward that star. The inset shows the region mapped in \HI\ self-absorption by \citet{McClure:2006}, 
while the blue rectangle outlines the area of the Pipe Nebula analyzed by \citet{Franco:2010}. 
Only three of the Na\,\textsc{i} sight lines intersect the \HI\ region studied in this work, 
underscoring that earlier distance estimates relied on a much broader area than the one considered here.
\label{fig:map_nai}}
\end{figure}

\begin{figure*}
\plotone{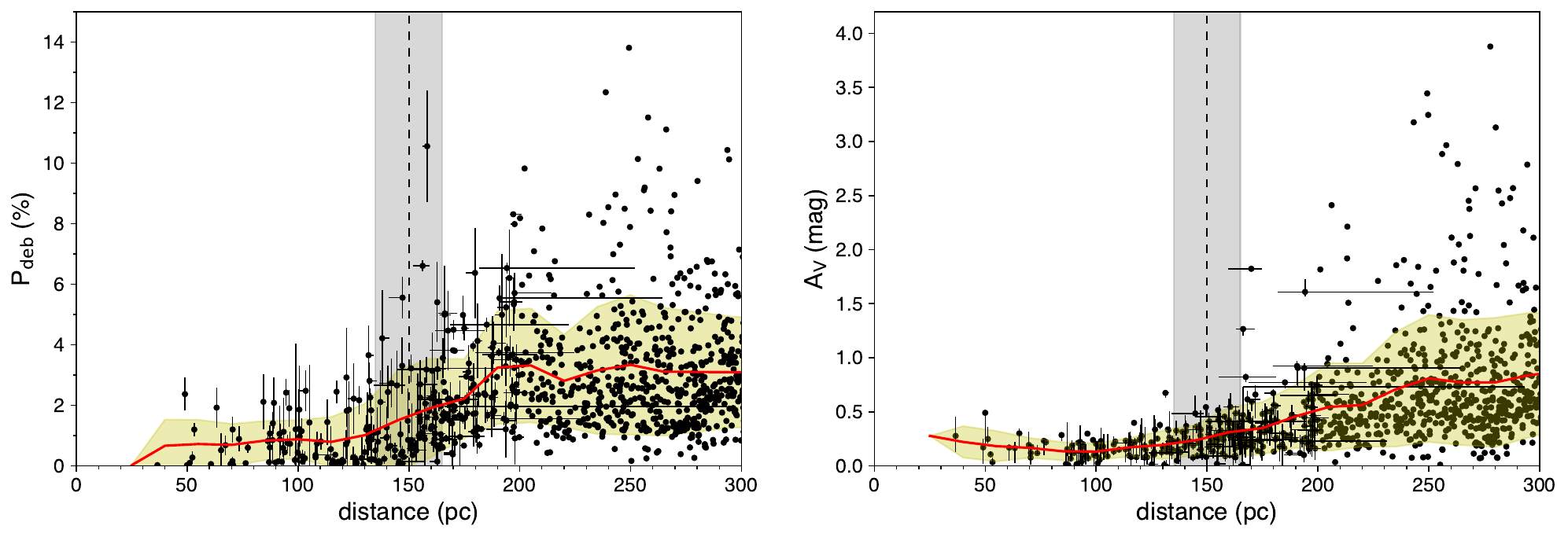}
\caption{Distance to the R--C cloud. 
\textit{Left:} debiased stellar polarization, $P_{\rm deb} = \sqrt{P^{2} - \sigma_{P}^{2}}$, as a 
function of stellar distance from \citet{Anders:2022}. The red curve shows the mean polarization in 
30 pc bins (stepped every 15 pc), while the shaded region represents the $\pm 1\sigma$ dispersion. 
The mean polarization remains low ($\langle P_{\rm deb} \rangle  \lesssim 1\%$) out to $\sim$130 pc, 
after which a clear rise is observed, reaching a plateau near $\sim$200 pc.  
\textit{Right:} visual extinction $A_{V}$ from \citet{Anders:2022} as a function of stellar distance, 
showing a similar increase beginning at $\sim$140--160 pc and saturating beyond $\sim$200 pc. 
The red curve and the shaded region have the same meaning as in the panel on the left.
In both panels, the vertical gray band marks the inferred distance to the R--C cloud ($150 \pm 15$ pc).  
Only stars within 200 pc are shown with individual uncertainties to maintain figure clarity.  
These parallel trends in polarization and extinction independently confirm the location of the R--C 
cloud and indicate that most of the intervening material lies at the $\sim$150 pc layer.
\label{fig:polxdist}}
\end{figure*}

Before analyzing the distance estimate to the R--C cloud in more detail, it is instructive to look into 
the spatial distribution of the stars, represented by the circles in Fig.\,\ref{fig:map_nai}, used by 
\citet{Crutcher:1984}. The size of the circles is proportional to the estimated interstellar Na\,{\sc i} 
column density. The small image superimposed on the figure is the area  mapped in \HI\ by 
\citet{McClure:2006}. The first point to be noticed is that the region covered by \citet{Crutcher:1984} 
is much larger than the area surveyed by \citet{McClure:2006}, and analyzed in this work. In fact, 
only three of the stars have line of sight passing through our region of interest, with two of them 
having minimum column density values. HD\,158704 is one of these stars and its \textit{Gaia}'s 
parallax ($\pi = 7.1748\pm0.078$ mas), places it at a distance of 139$\pm$2 pc. It is reasonable, 
then, to assume that the \HI\ cloud is located beyond this distance.  

Figure\,\ref{fig:polxdist} (left) shows the distribution of linear polarization as a function of the stellar 
distance from \citet{Anders:2022}. In order to have a representative sample of stars with low degree of 
polarization, no restrictions were imposed on the obtained $P/\sigma_P$, so the debiased degree of polarization 
($P_\mathrm{deb} = \sqrt{P^2 - \sigma_P^2}$) was plotted as a function of the distance. The red
curve represents the average polarization value estimated in bins of 30 pc in steps of 15 pc. The 
colored area denotes the $\pm 1\sigma$ interval. It is clear from this diagram that the mean 
observed polarization is almost constant ($\langle P_\mathrm{deb} \rangle \la1$ \%) up to a distance of about
130 pc, with several stars showing very low values of polarization degree in this distance interval,
beyond which an increasing trend is observed up to a distance of about 200 pc, becoming 
almost constant again, at a level of $P_\mathrm{deb} \approx 3$ \%.

The visual extinction profile, $A_V$ (Fig.\,\ref{fig:polxdist}, right), derived from \citet{Anders:2022}, increases at 
$\sim$140--160 pc, closely tracking the rise in debiased polarization (Fig.\,\ref{fig:polxdist}, left). Both quantities plateau
 beyond $\sim$200 pc, indicating that most of the intervening material lies within this distance range. Together with 
 the transition in Na\,{\sc i} absorption, these tracers consistently locate the dominant absorbing and polarizing material
  at $150 \pm 15$ pc.

This estimate improves upon the classical value of $125 \pm 25$ pc \citep{Crutcher:1984}, benefiting from \textit{Gaia} 
parallaxes and a larger stellar sample. The inferred distance is consistent with that of the Pipe Nebula 
\citep[e.g.,][]{Alves:2007, Dzib:2018}\footnote{It is worth noting that the distance obtained 
by \citet{Alves:2007} for the Pipe Nebula used stellar-polarization data and {\it Hipparcos}'s trigonometric 
parallaxes. A reanalysis of these data using \textit{Gaia}'s 
parallaxes pushes the Pipe Nebula a few parsecs away, in perfect agreement with the value suggested for 
the distance to the R--C cloud in this work. The same interval of distance was proposed by \citet{Dzib:2018} for 
{\it Barnard} 59, the densest part of this complex and the only one presenting evidence of star formation 
activity.}, whose line of sight overlaps the R--C cloud (Fig.\,\ref{fig:map_nai}). The alignment of the 
magnetic field in both regions \citep{Franco:2010} supports a physical association, in which the R--C cloud 
traces the diffuse atomic envelope of a larger magnetized complex that includes the Pipe Nebula. In this 
picture, the R--C filaments may represent magnetically guided flows feeding the molecular cloud, consistent 
with models of magnetically regulated cloud assembly.

\subsection{Background Structures}\label{subsec:back_clouds}

For the analysis of the magnetic field, it is important to assess the presence of additional structures along the line of sight. 
The distributions of visual extinction, polarization degree, and polarization angle as functions of stellar distance up to 
1.6~kpc are presented in Appendix~\ref{app:ext} (Figure~\ref{fig:pol_dist_full}). Individual measurements are shown 
as gray points. For visual extinction and polarization, red curves indicate running means computed in 100 pc bins 
(stepped by 50 pc), with shaded regions representing the corresponding $\pm1\sigma$ dispersion. For polarization 
angles, red symbols denote mean values with their standard deviations. The average extinction density (mag pc$^{-1}$) 
from \citet{Edenhofer:2024} is also shown for comparison.

Following the sharp increase at $\sim$150--200 pc, both extinction and polarization vary only gradually with distance. 
The mean $A_V$ remains $\lesssim 1$ mag between $\sim$400 and 900 pc, with little change in dispersion, and then 
increases more steadily beyond $\sim$900 pc, reaching $\gtrsim 2$ mag at $\sim$1.6 kpc. The polarization degree 
shows a similar behavior, rising to $\sim$3--4\% at $\sim$150--200 pc and increasing slowly thereafter, approaching 
$\sim$5\% at $\sim$1.6 kpc, while maintaining a significant dispersion at all distances.

The polarization angles exhibit a small shift beyond $\sim$200--300 pc and remain approximately constant over the 
next $\sim$700 pc, with only a modest change at distances $\gtrsim 1$ kpc, where the line of sight intersects the 
Sagittarius--Carina arm.

Overall, the modest variations in extinction and polarization beyond $\sim$200 pc and up to $\sim$1 kpc indicate 
the absence of significant additional dust structures along the line of sight behind the R--C cloud within this 
distance range.

\section{Comparison between Stellar and \textit{Planck} Polarizations}
\label{sec:planck_comparison}

To investigate the relationship between the magnetic field traced by optical starlight polarization 
and that inferred from thermal dust emission, we retrieved the $353$\,GHz Stokes $Q$ and $U$
maps together with their corresponding variances $QQ$ and $UU$ maps 
from the \textit{Planck} Legacy Archive (Appendix\,\ref{app:planck}).
Because plotting all individual stellar-polarization measurements (more than $9\times10^{4}$ stars 
with $P/\sigma_{P} \ge 5$) would produce an overcrowded and uninformative figure, we adopted a 
spatial averaging strategy described in Appendix \ref{app:ave}. 
This spatial averaging returned mean polarization and angles for 1044 cells.

\subsection{Comparison with \textit{Planck} Dust Polarization}\label{subsec:planck_dust}

\begin{figure*}[htb!]
\plotone{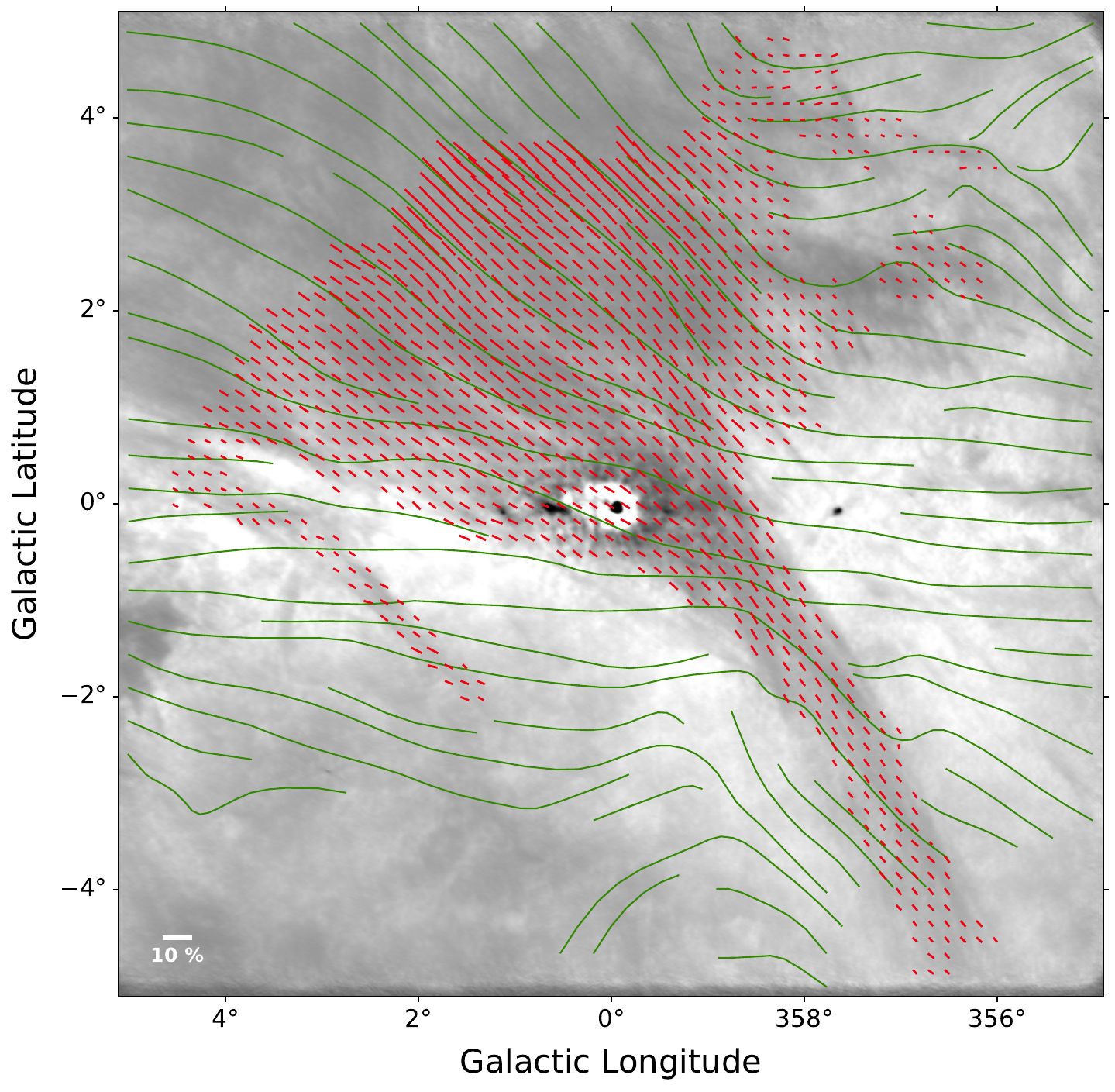}
\caption{Plane-of-sky magnetic-field orientations inferred from optical stellar polarimetry (red segments) 
and \textit{Planck} 353 GHz dust-emission polarization (green lines), overlaid on the \HI\ absorption map. 
Each stellar segment corresponds to the average Stokes parameters within $10' \times 10'$ spatial cells (see the text). 
The close agreement between the two tracers demonstrates that starlight and dust emission probe the same large-scale 
magnetic-field geometry across the R--C cloud. The segment scale for the optical polarimetry is provided in 
the lower-left corner.
\label{fig:map_pol}}
\end{figure*}

Figure~\ref{fig:map_pol} presents the plane-of-sky magnetic-field orientation derived from stellar 
polarimetry (red segments) and from \textit{Planck} 353\,GHz dust-emission data (green lines), 
overlaid on the \HI\ self-absorption map. 
The resulting magnetic-field patterns show remarkable agreement: both tracers reveal a coherent 
northeast-southwest orientation closely following the filamentary structure of the R--C cloud. 
This concordance demonstrates that starlight and dust emission are probing the same global 
magnetic-field geometry across the region. 

\begin{figure}[htb!]
\plotone{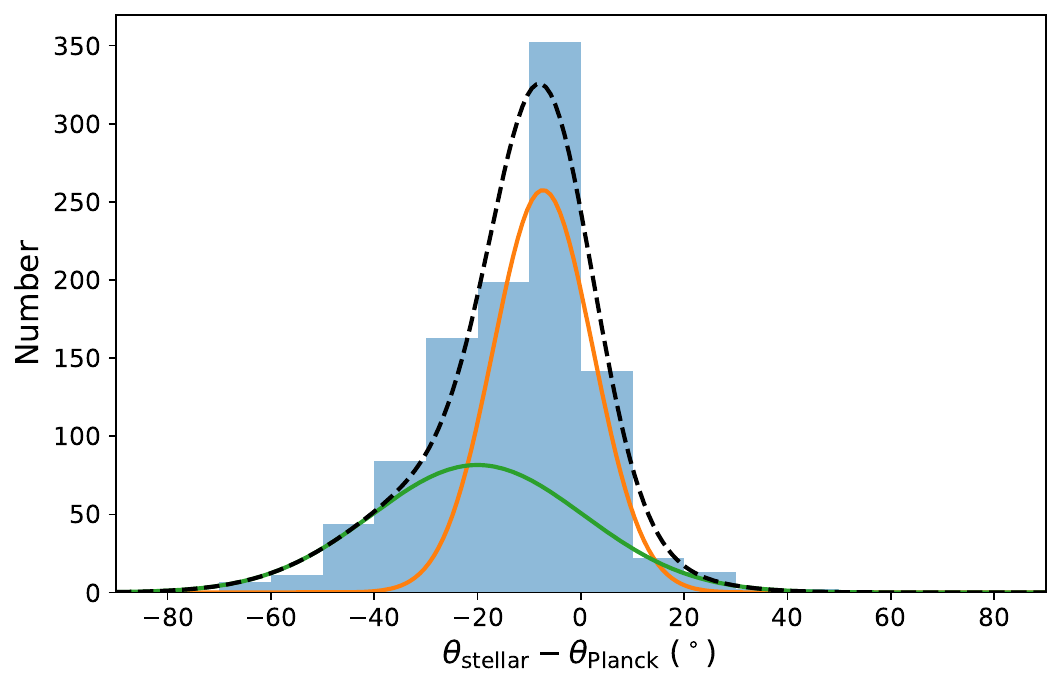}
\caption{Distribution of angular differences $\Delta\theta = \theta_{\rm stellar} - \theta_{\rm Planck}$ 
using $10^{\circ}$ bins, shown with a two-component Gaussian mixture model. Individual Gaussian 
components and the combined model are overplotted. The distribution peaks near 
$\Delta\theta \sim -7\degr$, indicating strong alignment between the two independent tracers 
of the POS magnetic field.
\label{fig:norm_hist}}
\end{figure}

To quantify the agreement, we computed the angular offset between the two tracers,

\begin{equation}
    \Delta\theta \equiv \theta_{\rm stellar} - \theta_{\rm Planck},
\end{equation}
where $\theta_{\rm Planck}$ and $\theta_{\rm stellar}$ are the average angles defined in 
Appendices \ref{app:planck} and \ref{app:ave}.
The obtained distribution is presented in  Fig.~\ref{fig:norm_hist} and shows a strong central 
concentration around  $\Delta\theta \sim -10\degr$, along with an extended negative tail and 
a smaller number of positive outliers. Basic statistical indicators highlight the asymmetry: 
the mean and median of the distribution are $-12\fdg45$ and $-9\fdg39$, respectively, 
with a skewness of $-0.21$ and a kurtosis of $2.24$. 

A single symmetric Gaussian does not adequately represent the distribution. A two-component 
Gaussian mixture model provides a much better description, revealing:
\begin{eqnarray}
     \mu_1 = -7\fdg3, & \sigma_1 = 9\fdg7, &  w_1 = 0.60, \nonumber \\ 
    \mu_2 = -20\fdg0, &  \sigma_2 = 20\fdg6,  &  w_2 = 0.40, \nonumber
\end{eqnarray}
where $w_i$ denotes the weight of each component. 

It is instructive to compare these dispersions with the expected uncertainty from the 
measurements. Considering typical uncertainties of about $3^\circ$ for our mean stellar-polarization angles and 
$12^\circ$ for \textit{Planck}, the combined uncertainty should be on the order of
$\sigma_{\rm exp} \approx \sqrt{3^2 + 12^2} \approx 12\fdg4$. The narrower component, with 
$\sigma \approx 9\fdg7$, is therefore consistent with (and even slightly smaller than) the expected observational 
scatter. This suggests that this component likely represents regions where the magnetic-field orientation is 
coherent along the line of sight, and where the difference between stellar and \textit{Planck} measurements 
is dominated primarily by measurement uncertainties. In these regions, both tracers are effectively probing 
the same underlying magnetic-field geometry.

In contrast, the broader lower-amplitude component, with $\sigma \approx 20\fdg6$, significantly exceeds 
the expected uncertainty,  and accounts for the pronounced negative tail. As seen in Fig.~\ref{fig:map_pol}, 
most of the large offsets originate from the region $| b | \le 2\degr$, where \textit{Planck} 
emission likely includes a significant background contribution from the Galactic plane. 

The systematic increase in the angular offset between the stellar and \textit{Planck}
magnetic-field orientations toward low Galactic latitudes is expected given
the very different volumes sampled by the two tracers. Optical stellar
polarization probes the magnetic field only up to the distance of each star
and is therefore primarily sensitive to the local interstellar medium,
whereas \textit{Planck} 353\,GHz dust-emission polarization integrates the signal over
the entire line of sight through the Galactic disk. Near the Galactic plane,
this line-of-sight integration includes contributions from multiple spiral
arms and large-scale disk structures, where the ordered Galactic magnetic
field is known to lie predominantly parallel to the plane
\citep[e.g.,][]{Han:2006, Beck:2013, Haverkorn:2015, Planck_XIX:2015}.
As a result, \textit{Planck} polarization angles at low $|b|$ reflect a weighted
average of magnetic-field orientations over kiloparsec scales, while the
stellar-polarization angles trace primarily the component of the magnetic field in the 
plane of sky that acts on the nearby R--C cloud at
$\sim$150\,pc. This geometric mismatch naturally leads to larger angular
offsets close to the Galactic plane and does not imply a physical
misalignment within the cloud itself, but rather the increasing influence of
unrelated foreground and background dust along the line of sight.

\section{Discussion}\label{sec:discussion}

The R--C cloud exhibits the same magnetically regulated, filamentary morphology identified 
in large-scale high-latitude \HI\ surveys, where CNM structures are preferentially organized 
into magnetic field–aligned filaments \citep[e.g.,][]{Clark:2014, Kalberla:2020, Lei:2023}, 
reinforcing its classification as part of a broader class of magnetically aligned CNM structures.

\begin{figure}[htb!]
\plotone{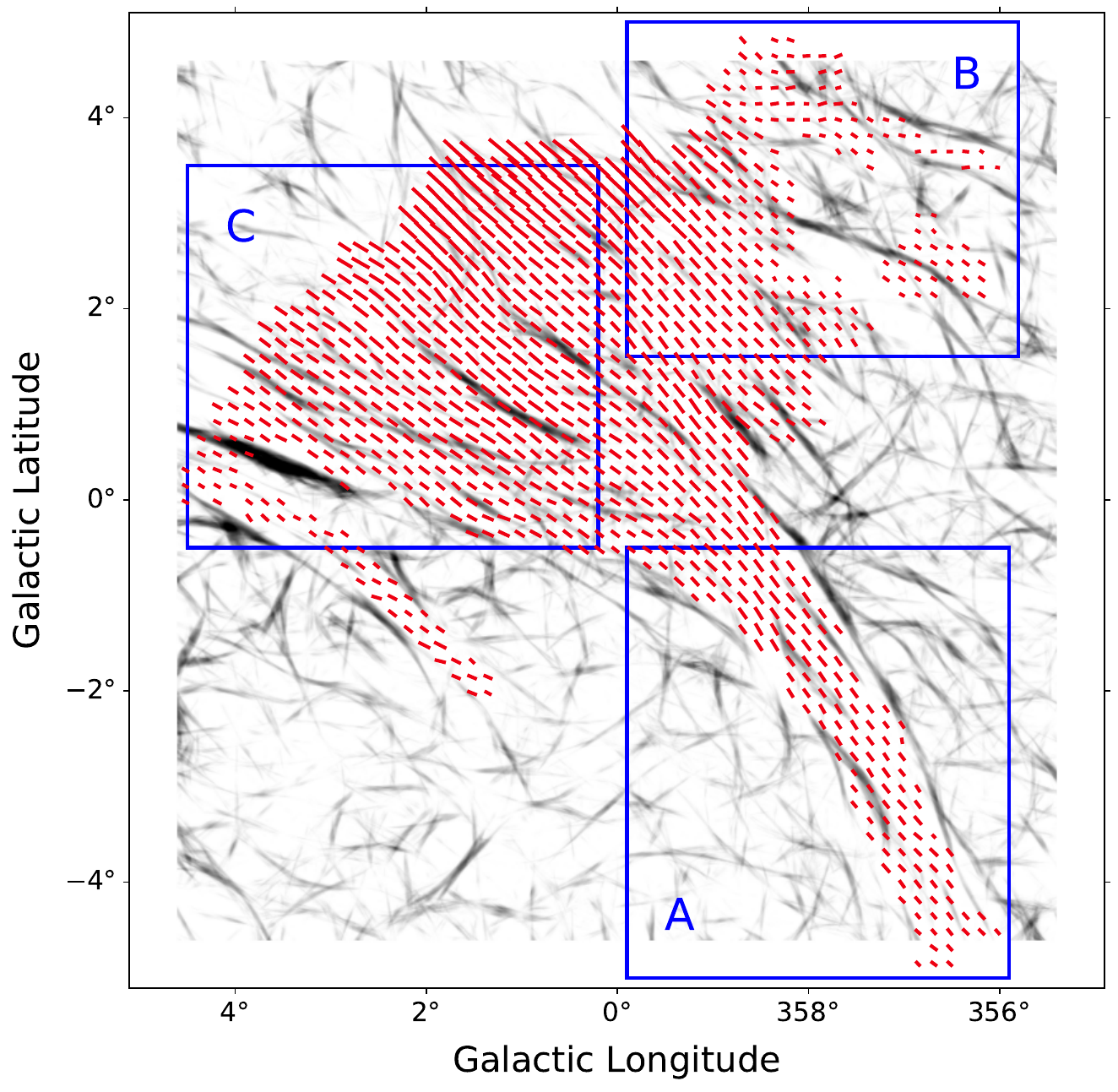}
\caption{Comparison between the \HI\ filament orientations (black linear structures), extracted using 
the RHT method \citep{Clark:2014}, and the magnetic-field direction derived from stellar polarization (red segments). 
The strong parallelism between the RHT-extracted \HI\ filaments and the stellar magnetic-field 
directions provides independent confirmation that the CNM structures in the R--C cloud are magnetically 
aligned. This agreement reinforces the interpretation of the R--C cloud as a magnetically dominated 
environment.
\label{fig:map_fil}}
\end{figure}

\begin{figure*}[h!]
\gridline{\fig{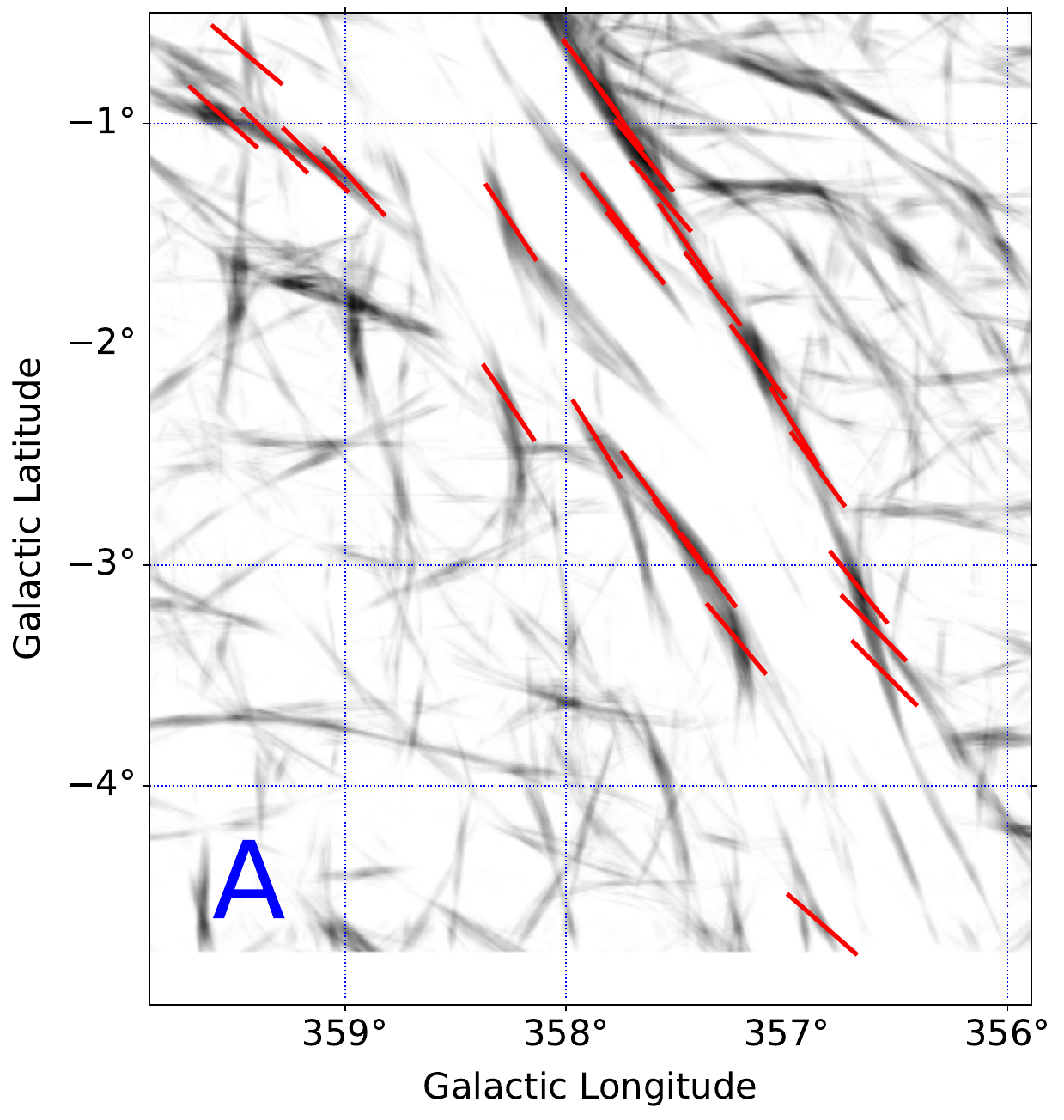}{0.48\textwidth}{}
          \fig{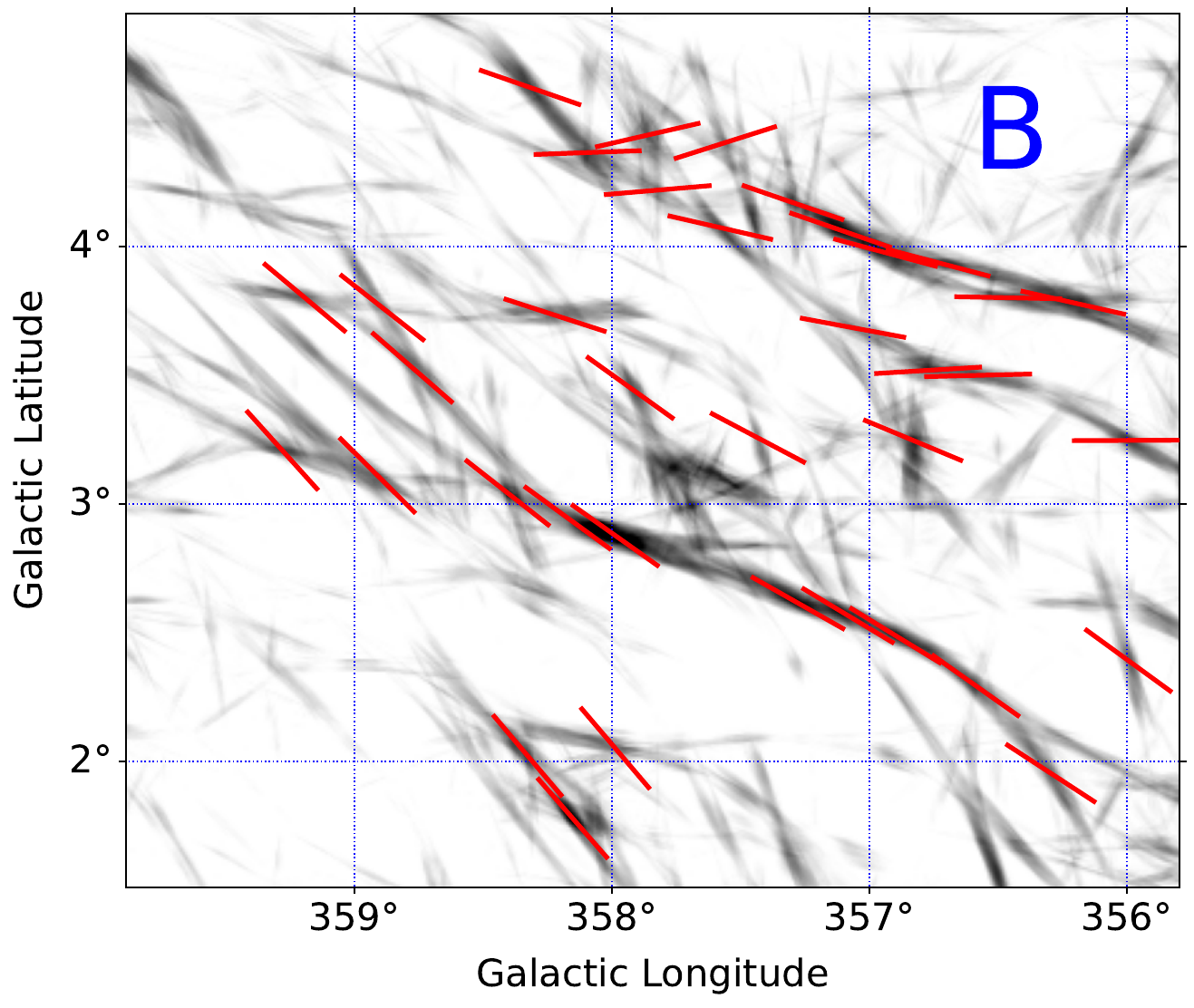}{0.48\textwidth}{}
          }
\gridline{\fig{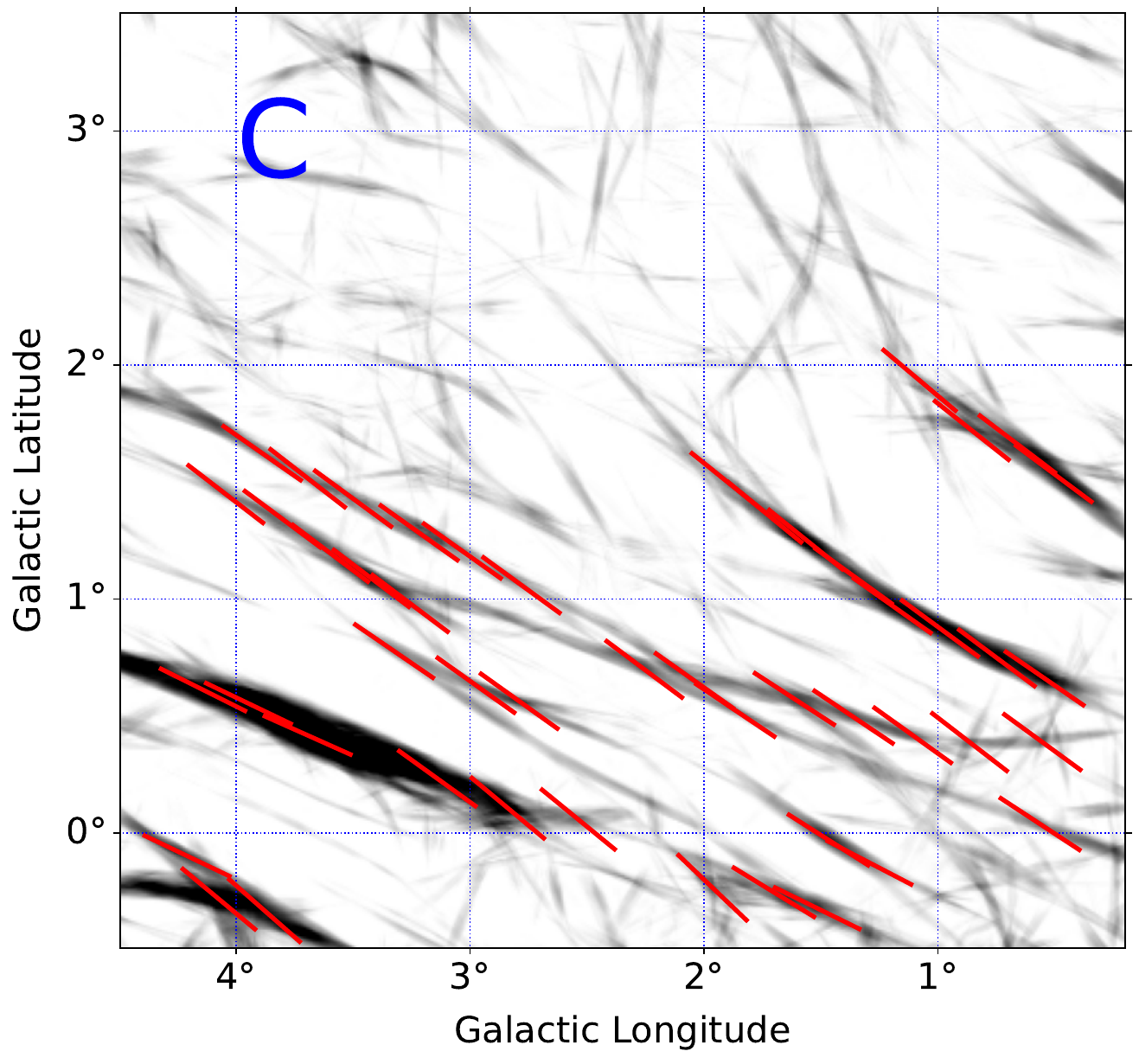}{0.48\textwidth}{}
          }
\caption{Zoomed-in views of three representative subregions (corresponding to Fields A, B, and C in 
Fig.\,\ref{fig:map_fil}) showing \HI\ self-absorption filaments extracted with the RHT analysis 
(in grayscale) overlaid with plane-of-sky magnetic-field orientations derived from stellar 
polarization (red segments) and positioned along the filaments using spatial sampling without a 
fixed grid. The filamentary structures closely follow the polarization segments that trace the projected
magnetic-field orientation.
}\label{fig:fields_detail}
\end{figure*}

Figures\,\ref{fig:map_fil} (full field) and \ref{fig:fields_detail} (details) compare the orientation 
of \HI\ filaments, extracted using the Rolling Hough Transform (RHT; \citealt{Clark:2014} -- see 
Appendix\,\ref{app:rht}), with the POS magnetic-field direction derived from stellar polarization. 
The strong parallelism between the RHT-identified filaments and the polarization segments provides 
compelling evidence that the \HI\ structures are not randomly oriented nor primarily shaped by turbulence. 
Instead, this alignment is generally interpreted as arising from magnetized turbulence and is 
consistent with a sub-Alfv\'enic regime, in which the magnetic field introduces anisotropy in the 
turbulent cascade, limiting motions perpendicular to the field and favoring gas flows along field lines 
\citep{Soler:2013, Chen:2016}.

To quantify the relative orientation between the \HI\ filamentary structures and the magnetic 
field, filament positions were identified from the intensity map and their orientations compared 
with the magnetic-field direction inferred from nearby stellar-polarization measurements. At each 
chosen filament location, a mean polarization angle was estimated using stars within a fixed angular 
neighborhood, following the procedure described in Appendix\,\ref{app:ave}, and considering only 
high signal-to-noise measurements ($P/\sigma_{P} \geq 5$).

\begin{figure}
\plotone{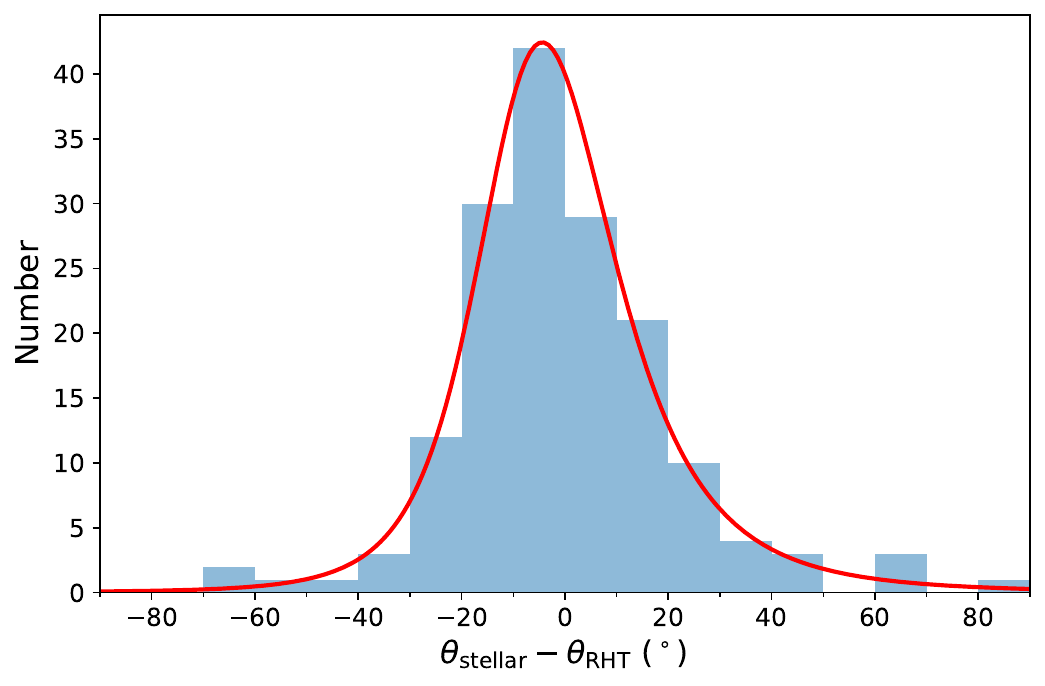}
\caption{Distribution of the angular differences $\Delta\theta = \theta_{\mathrm{stellar}} - \theta_{\mathrm{RHT}}$. 
The histogram shows the observed frequency of angle offsets between the stellar polarization and the RHT-derived 
orientations. The red solid line represents the best-fitting skew-$t$ distribution, which provides an excellent description 
of the asymmetric shape and extended tails of the data. The distribution is strongly peaked near zero, indicating a 
high degree of alignment between the two tracers, with a slight skewness reflecting residual systematic deviations.}
\label{fig:theta_rht}
\end{figure}

The resulting distribution of angular differences, 
$\Delta\theta = \theta_{\mathrm{stellar}} - \theta_{\mathrm{RHT}}$, is shown in 
Fig.\,\ref{fig:theta_rht}. The distribution is strongly peaked near $0^\circ$, indicating a 
preferred alignment between \HI\ filaments and the magnetic field, but it also exhibits extended 
heavy tails and a moderate positive skewness. These features reveal that, while alignment is 
globally maintained, significant local deviations occur.

To model this behavior, we compared a single Gaussian, a Gaussian mixture model (GMM), and a 
skewed Student's $t$ (skew-$t$) distribution. The Gaussian model fails to reproduce both the sharp 
central peak and the extended tails, while the GMM provides a better fit at the cost of introducing 
multiple components without clear physical interpretation. In contrast, the skew-$t$ distribution 
offers a parsimonious and accurate description of the data. The best-fit parameters 
($\nu \approx 3.33$, $\alpha \approx 0.77$) indicate strong non-Gaussianity and intrinsic 
asymmetry, supporting a scenario in which the magnetic alignment is perturbed by turbulence, 
line-of-sight superposition, and local dynamical processes.

The heavy-tailed nature of the distribution implies that large angular deviations occur more 
frequently than expected from Gaussian statistics, suggesting the influence of intermittent or 
localized processes such as shocks, shear flows, or anisotropic turbulence. The observed 
positive skewness further indicates that positive angular deviations are either more frequent or 
more extreme than negative ones. This asymmetry may arise from a combination of projection effects, 
biases in the RHT detection algorithm, or intrinsic physical mechanisms that systematically distort 
the magnetic field, such as compressive flows.

\begin{deluxetable*}{cccccccc}
\tablenum{1}
\tablecaption{Magnetic-field properties of the Riegel–Crutcher cloud 
    derived from angular dispersion analysis.}
\tablewidth{0pt}
\tablehead{
\colhead{Region} & \colhead{N} & \colhead{$b$} & \colhead{$\langle B_t^2 \rangle^{1/2}/B_0$} &
\colhead{$\langle A_V \rangle$} & \colhead{$n_H$} & \colhead{$\rho$} & \colhead{$B_\mathrm{pos}$} \\
\nocolhead{} & \nocolhead{} & \colhead{(rad)} & \nocolhead{} & \colhead{(mag)} & \colhead{ (cm$^{-3}$)} &
\colhead{($10^{-22}$ g cm$^{-3}$)} & \colhead{($\mu$G)}
}
\startdata
Field A & 2360 & 0.119 $\pm$ 0.012 & 0.085 $\pm$ 0.008 & 0.853 $\pm$ 0.380 & 153 $\pm$ 69 & 3.58 $ \pm$ 1.63 & 55 $\pm$ 15 \\
Field B & 3360 & 0.136 $\pm$ 0.005 & 0.097 $\pm$ 0.003 & 1.195 $\pm$ 0.557 & 216 $\pm$ 101 & 5.05 $\pm$ 2.37 & 56 $\pm$ 15 \\
Field C & 6711 & 0.113 $\pm$ 0.008 & 0.080 $\pm$ 0.006 & 1.351 $\pm$ 0.660 & 244 $\pm$ 118 & 5.72 $\pm$ 2.76 & 72 $\pm$ 20 \\
\enddata
\tablecomments{In the second column, N represents the number of stars used in the ADF calculation.}
\end{deluxetable*}\label{tab:adf}

A more detailed view of the magnetic-field structure is provided by the angular dispersion 
analysis. Previous estimates of the magnetic-field strength in the R--C cloud span a wide range, 
from $\sim$19~$\mu$G to $\sim$60~$\mu$G \citep{McClure:2006, Clark:2014}, reflecting both 
observational limitations and methodological assumptions. These estimates rely on the 
Davis--Chandrasekhar--Fermi (DCF) method \citep{Davis:1951, Chandrasekhar:1953}, which relates the 
magnetic-field strength to the dispersion of polarization angles, the velocity dispersion, and the 
gas density.

For example, \citet{McClure:2006} adopted $\rho = 1.1 \times 10^{-21}$ g cm$^{-3}$ and 
$\sigma_v = 1.4$ km s$^{-1}$, and used polarization data from \citet{Heiles:2000} for only 56 stars, 
obtaining $\langle B \rangle \simeq 60~\mu$G. However, this estimate is limited by the small sample 
size and by the assumption of homogeneous density and isotropic turbulence, which are unlikely to 
hold in a structured medium. Alternatively, \citet{Clark:2014} inferred field strengths of 
$\sim$19--50~$\mu$G by assuming that RHT filaments trace the magnetic-field direction, although this 
approach depends critically on the validity of that assumption.

The present polarization survey enables a more robust determination of the magnetic properties. 
Using the angular dispersion function (ADF) method (Appendix\,\ref{app:adf}), we estimate the 
magnetic-field strength in three subregions (the Fields A, B, and C in Fig.\,\ref{fig:map_fil}). The results, 
summarized in Table\,\ref{tab:adf}, reveal distinct magnetic regimes across the cloud.

Fields A and C are characterized by relatively ordered magnetic fields, as indicated by their 
low angular dispersions and small turbulent-to-ordered field ratios. In these regions, the ADF 
shows a rapid rise at small separations followed by a flattening, consistent with a short 
turbulent correlation length. In contrast, Field B exhibits the largest angular dispersion and 
turbulent fraction, with an ADF that increases smoothly over larger scales, suggesting a more 
disordered field geometry possibly influenced by large-scale gradients or multiple structures 
along the line of sight.

These differences are also reflected morphologically in Fig.\,\ref{fig:fields_detail}. Fields A and C 
show a clear alignment between polarization segments and \HI\ filaments, indicating a strong coupling 
between the magnetic field and the gas. Field B, on the other hand, displays a weaker correspondence, 
with polarization angles showing larger scatter and frequent deviations from filament orientations, 
consistent with enhanced turbulence and reduced magnetic coherence.

The derived POS  magnetic-field strengths range from $\sim$55~$\mu$G to $\sim$72~$\mu$G, 
slightly higher than previous estimates. This difference is primarily driven by the smaller angular dispersions 
obtained with the ADF method ($b = 0.113$--0.136 rad), compared to $\sigma_\theta \simeq 0.192$ rad in 
\citet{McClure:2006}. Given that $B \propto 1/\sigma_\theta$, this reduction alone naturally leads to larger 
inferred field strengths.

On the other hand, we adopt lower mass densities than those assumed by \citet{McClure:2006}, which 
would act in the opposite direction, yielding smaller magnetic-field strengths. This choice is directly tied 
to the assumed line-of-sight thickness of the medium: a larger thickness implies a lower volume density 
for a fixed column density. Therefore, while our density assumptions tend to decrease the estimated field 
strengths, this effect is outweighed by the smaller angular dispersions derived with the ADF method. These 
combined factors indicate that the higher values reported here are mainly controlled by the improved 
characterization of the magnetic-field fluctuations, while still being subject to systematic uncertainties 
associated with the poorly constrained cloud thickness.

The inferred magnetic-field strengths in the three regions imply Alfvén speeds significantly higher than 
the observed nonthermal velocity dispersion ($\sigma_v \approx 1.4$ km s$^{-1}$). As a consequence, 
the corresponding Alfvén Mach numbers are well below unity ($M_\mathrm{A} \ll 1$), indicating a clearly sub-Alfvénic 
regime. In this limit, the magnetic field plays a dynamically dominant role, regulating the gas motions and 
reducing the impact of turbulent fluctuations. The resulting dynamics are therefore highly anisotropic, with 
gas motions preferentially guided along the magnetic-field lines and strongly inhibited in the perpendicular 
direction. This interpretation is fully consistent with the observed magnetic-field morphology, the strong 
alignment between \HI\ filaments and polarization angles, and the high degree of coherence in 
Fields A and C. Even when accounting for uncertainties in density and cloud thickness, the results 
consistently indicate that the R--C cloud is a magnetically dominated environment.

Taken together, the agreement between optical polarimetry, \HI\ morphology, and angular dispersion 
analysis provides a coherent and self-consistent picture in which magnetic fields play a central role 
in structuring the cloud. This consistency across independent diagnostics strengthens the reliability 
of the inferred field geometry and supports its dynamical importance.

\begin{figure*}[h!]
\plotone{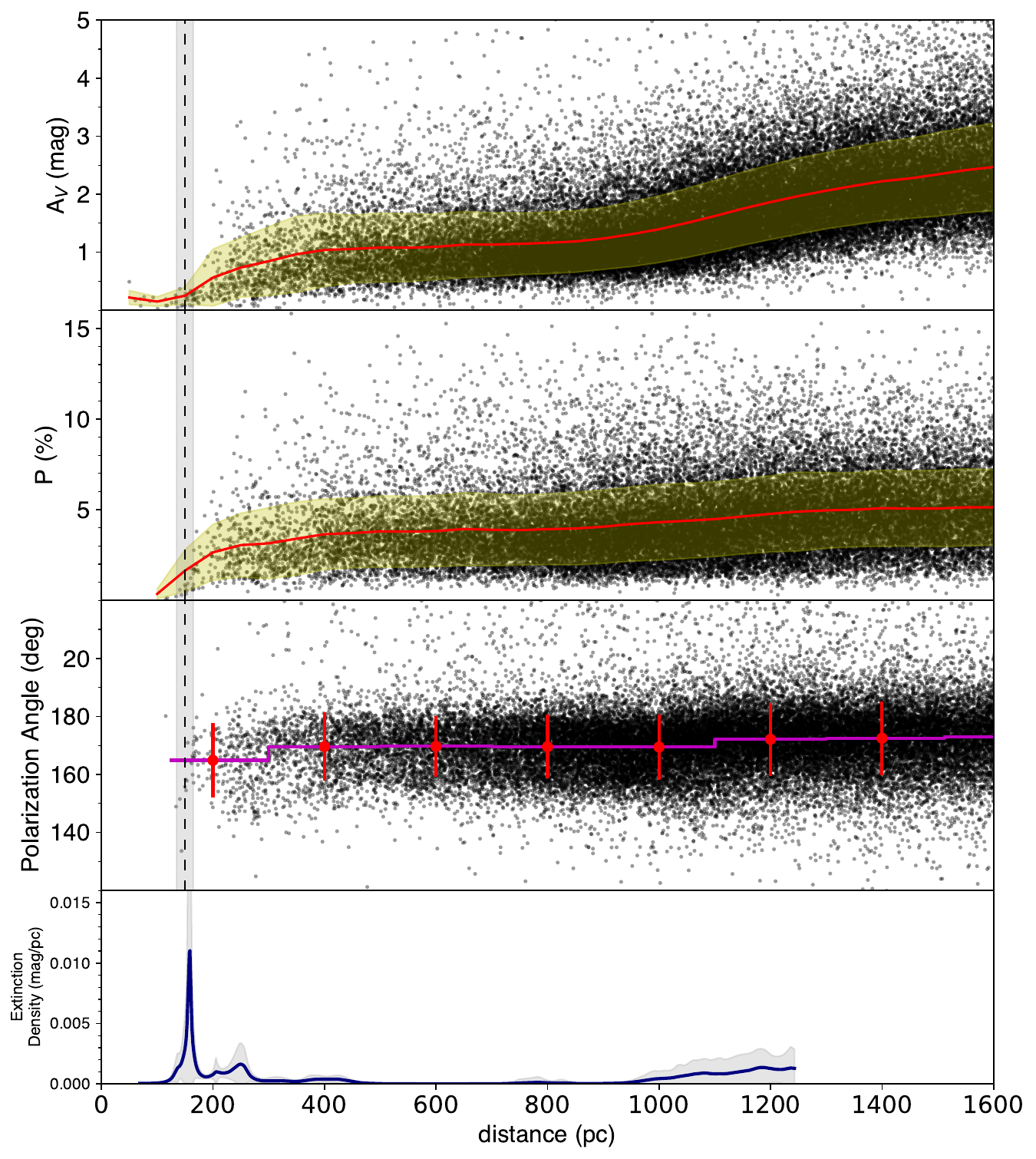}
\caption{
Distance dependence of extinction $A_V$, polarization $P$, and polarization angle $\theta$. The extinction 
and distances are taken from the STARHORSE  catalog \citep{Anders:2022}. 
Black points represent individual measurements; for the polarization panels, only sources with $P/\sigma_P \ge 5$
are included. The red curves indicate the mean trends as a function of distance, and the shaded bands 
correspond to the $\pm 1 \sigma$ dispersion. In the polarization angle panel, the red points show the binned 
mean values with their associated uncertainties, while the magenta curve traces the overall angular trend.
The vertical dashed line marks the characteristic distance of 150 pc estimated in this work, with the gray shaded 
region indicating its associated uncertainty. The bottom panel shows the extinction density profile from 
\citet{Edenhofer:2024}, where the gray shading denotes the uncertainty in the reconstructed density.
\label{fig:pol_dist_full}}
\end{figure*}

\begin{figure*}[hbt]
\gridline{\fig{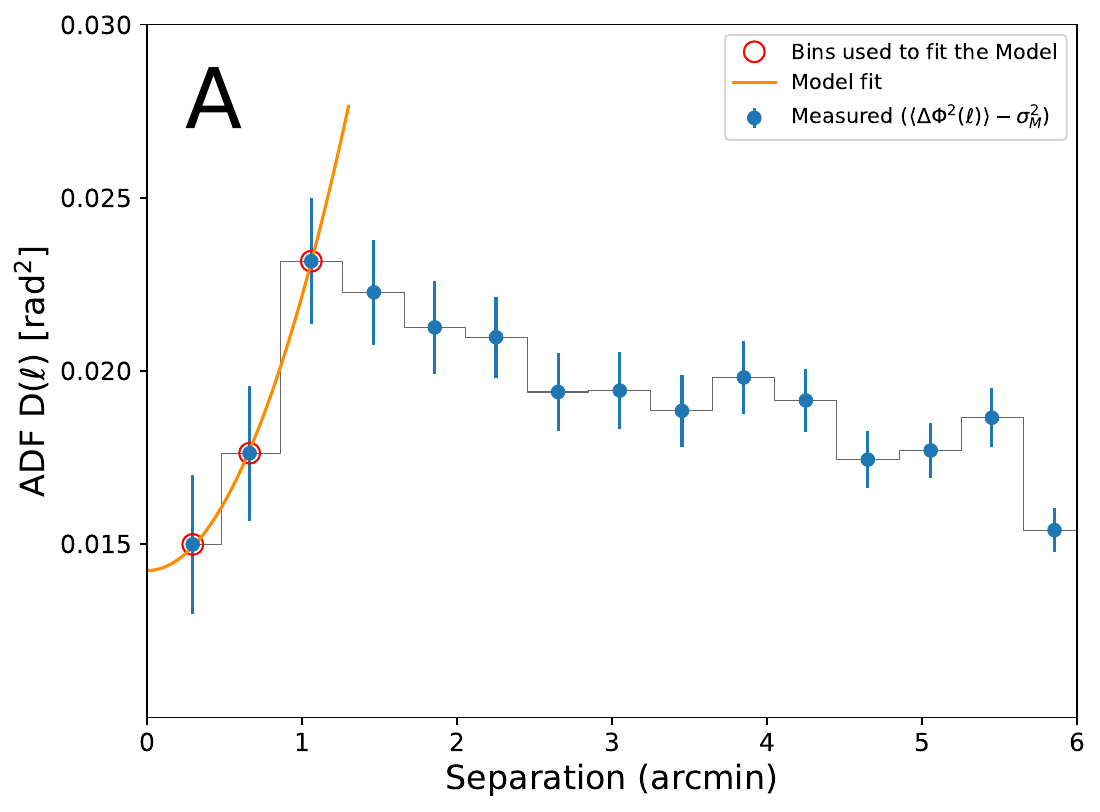}{0.48\textwidth}{}
          \fig{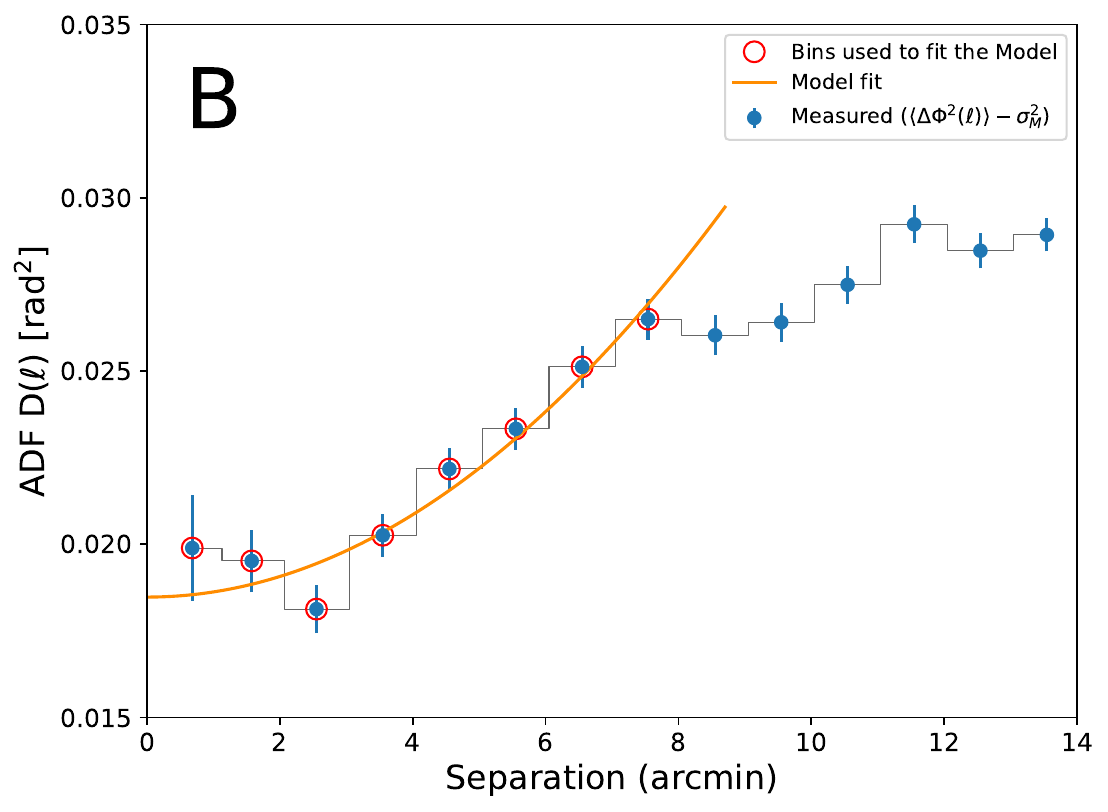}{0.48\textwidth}{}
          }
\gridline{\fig{adf_plot_field_C.pdf}{0.48\textwidth}{}
          }
\caption{Measured angular dispersion function, $( \Delta\Phi^2(\ell) - \sigma^2_\mathrm{M})$, as a function of angular 
separation $\ell$ for the three regions defined in Fig.\,\ref{fig:map_fil} (Fields A, B, and C). Blue points 
show the measured ADF, with vertical bars indicating statistical uncertainties. Orange curves represent the
best-fitting ADF models, and red symbols mark the bins used in the fits. Fields~A and~C are characterized by 
relatively ordered magnetic fields, but they differ in field strength and internal complexity. On the other hand, 
Field~B is dominated by a higher level of disorder and large-scale variations. The zero-separation intercept 
of the fitted models yields the turbulent angular dispersion parameter, $b$, used to estimate the 
plane-of-sky magnetic-field strength. At the estimated distance to the R--C cloud, 1~arcmin corresponds to 
$\sim$0.04~pc. 
}\label{fig:adf_plots}
\end{figure*}

\section{Conclusions}

We have presented a comprehensive optical polarimetric study of the Riegel--Crutcher cloud, combining a large 
dataset of more than $9\times10^{4}$ high signal-to-noise stellar-polarization measurements with 
\textit{Gaia} DR3 data. This dataset enables a detailed characterization of the plane-of-sky magnetic 
field and its relation to the filamentary structure of the cold neutral medium.

Our main results can be summarized as follows:

\begin{enumerate}

\item Using a combination of Na\,{\sc i} absorption, stellar polarization, and visual extinction, we constrain 
the distance to the R--C cloud to $150 \pm 15$ pc. These independent tracers consistently indicate that the 
bulk of the absorbing and polarizing material is concentrated within a relatively narrow distance range.

\item The magnetic-field orientation inferred from optical stellar polarimetry is in excellent agreement with 
that derived from \textit{Planck} 353\,GHz dust-emission polarization. The angular-offset distribution peaks 
near $\Delta\theta \sim -7^\circ$, with a dispersion consistent with a predominantly coherent magnetic field.

\item The dispersion in the angular differences between the two tracers cannot be fully explained by measurement 
uncertainties alone. Instead, it reflects a combination of intrinsic magnetic-field variations and line-of-sight integration 
effects, particularly toward low Galactic latitudes.

\item The close alignment between the magnetic-field orientation and the filamentary structure of the R--C cloud 
provides strong evidence that magnetic fields play a key role in shaping the morphology of the cold atomic gas.

\item Analysis of the angular dispersion function yields plane-of-sky magnetic-field strengths of  
$B_\mathrm{pos} \approx 55 \pm 15$~$\mu$G, $56 \pm 15$~$\mu$G, and $72 \pm 20$~$\mu$G across 
the three regions, suggesting a magnetically dominated, sub-Alfv\'enic cold neutral medium across the probed 
scales. Although these absolute field strengths are subject to assumptions about cloud thickness, the finding 
that magnetic energy outweighs turbulent motions remains robust.

\item The overall picture that emerges is that of a magnetically organized structure, in which the large-scale field 
governs the global geometry of the cloud, while smaller-scale fluctuations---likely associated with 
turbulence---introduce localized variations.

\end{enumerate}

Taken together, these results underscore the key role of magnetic fields in structuring the diffuse 
interstellar medium, even in predominantly atomic environments. The R--C cloud provides a compelling 
example of how magnetic fields can imprint coherent, large-scale organization on cold gas, offering 
valuable constraints for theoretical models of ISM structure formation. However, translating these 
insights into robust quantitative estimates---particularly magnetic-field strengths derived via the 
Davis--Chandrasekhar--Fermi (DCF) method---remains fundamentally limited by uncertainties in 
local gas density. Since such uncertainties propagate directly into the inferred Alfv\'en speed and 
magnetic stresses, improving density determinations emerges as a critical step toward reducing 
systematic errors. Continued observational and theoretical efforts in this direction will therefore be 
essential for advancing a quantitative understanding of the Galactic magnetic field.

\begin{acknowledgments}
We sincerely thank the referee for their thorough and insightful review, which greatly 
contributed to improving the quality and clarity of this work. The authors also thank the 
staff of OPD/LNA (Brazil) for their hospitality and invaluable help during our observing runs. 
Z.Y.L. is supported in part by NASA 80NSSC20K0533, NSF AST-2307199, and 
JWST-GO-08872.003-A. This work used data from the \textit{Gaia} mission of the European 
Space Agency (ESA) (\url{https://www.cosmos.esa.int/gaia}), processed by 
the \textit{Gaia} Data Processing and Analysis Consortium (DPAC,
\url{https://www.cosmos.esa.int/web/gaia/dpac/consortium}). Funding for the DPAC
has been provided by national institutions, in particular the institutions
participating in the \textit{Gaia} multilateral agreement. We also used  data retrieved 
from the Planck Legacy Archive (\url{http://pla.esac.esa.int}), an ESA science mission with 
instruments and contributions directly funded by ESA Member States, NASA, and Canada.
\end{acknowledgments}

\vspace{5mm}
\facility{LNA:BC0.6m (LNA, Brazil).}

The stellar polarimetric data used for the analysis conducted in this work are available from the 
corresponding author upon reasonable request.

\software{
{\tt matplotlib} \citep{Hunter:2007}, {\tt astropy} \citep{Astropy:2013, Astropy:2018}, 
{\tt SOLVEPOL} \citep{Ramirez:2017}, {\tt TOPCAT} \citep{Taylor:2005}.}

\appendix

\section{Cross-match with the \textit{Gaia} Catalog}\label{app:cross}

We cross-matched our sources with the \textit{Gaia} catalog using a $1\arcsec$ 
search radius. Approximately 94\% of the sources have a unique counterpart within this radius. 
For sources with multiple candidates, we applied the likelihood-ratio (LR) method 
\citep{Sutherland:1992} to select the most probable match.

The positional term was modeled as a Gaussian in the angular separation $r$,
\begin{equation}
f(r) = \exp\left(-\frac{r^2}{2\sigma_\mathrm{pos}^2}\right),
\end{equation}
with $\sigma_\mathrm{pos} = 0\farcs3$.

We included a photometric term based on the difference between the observed \textit{Gaia} 
$B_P$ magnitude and that predicted from $V$ via a linear fit,
\begin{equation}
q(m) = \exp\left(-\frac{\Delta m^2}{2\sigma_\mathrm{mag}^2}\right),
\end{equation}
where $\sigma_\mathrm{mag} = 0\fm5$.

The background magnitude distribution, $n(m)$, was estimated empirically from the $B_P$ 
distribution of all candidates. The likelihood ratio is then
\begin{equation}
LR = \frac{q(m),f(r)}{n(m)}.
\end{equation}

The reliability (posterior probability) of candidate $i$ is
\begin{equation}
P_i = \frac{LR_i}{\sum_j LR_j + 1 - Q},
\end{equation}
where the sum is over all candidates for a given source and $Q$ is the prior probability 
that a true counterpart exists (we adopted $Q = 0.95$).

For multiple candidates, the source with the highest $P_i$ value was chosen as the counterpart, 
and the others were rejected.

\section{Distribution of the interstellar extinction and polarization}\label{app:ext}

The distance dependence of extinction, polarization, and polarization angle reveals a coherent 
interstellar structure that is broadly consistent with the profile inferred by \citet{Edenhofer:2024}. Both ($A_V$) 
and ($P$) exhibit a pronounced rise at ($d \le 200$) pc, coincident with the sharp peak 
in the \citet{Edenhofer:2024} distribution, indicating a dominant dust layer along the line of sight. Beyond this 
distance, ($A_V$) increases gradually while ($P$) shows a slower, cumulative growth, suggesting 
additional diffuse contributions not fully captured by the model. In contrast, the polarization angle 
stabilizes at ($\theta \sim 165\degr$) beyond the same distance, with relatively low dispersion, 
implying that the magnetic-field orientation is largely established within the main structure and 
remains coherent at larger distances. While the  \citet{Edenhofer:2024}  curve successfully reproduces the location 
of the primary dust concentration, the observed scatter and extended tails in all three parameters 
point to a more complex, multilayered interstellar medium with significant line-of-sight variation.

Figure\,\ref{fig:pol_dist_full} compares the behavior of the cumulative extinction and polarization with the differential 
extinction density averaged over the area surveyed in Fig.\,\ref{fig:map_pol}  ($| \ell | \le 5\degr,  | b | \le 5\degr$). 
In the top panel, the sharp peak in the extinction density at $\sim$150-200 
pc is clearly mirrored by a rapid rise in $A_V$, indicating the presence of a nearby dust structure that dominates 
the extinction budget at short distances. Beyond this feature, the extinction density decreases and then gradually 
increases again toward distances $\lesssim 1$ kpc, which is reflected in the steady, approximately linear growth of 
$A_V$. This correspondence is expected, as $A_V$ represents the integral of the extinction density along the 
line of sight.

In contrast, the polarization behavior (bottom panel) does not follow the same trend at larger distances. Although $P$ increases 
rapidly at short distances, coincident with the nearby extinction feature, it shows only a modest and smooth rise beyond 
$\sim$300 pc and does not exhibit a clear response to the renewed increase in extinction density around $\sim$1 kpc.
This apparent lack of correlation suggests that the additional dust contributing to the extinction at these distances is 
comparatively inefficient at producing polarized emission. Several factors may contribute to this effect. One possibility is that 
the magnetic-field geometry becomes more complex or tangled at larger distances, leading to partial cancellation of 
polarization along the line of sight. Alternatively, variations in grain alignment efficiency or changes in dust properties 
could reduce the polarization per unit extinction. Depolarization due to multiple overlapping structures with differing 
field orientations may also play a significant role. The combined evidence therefore indicates that, while extinction 
continues to accumulate with distance, the polarization signal becomes increasingly insensitive to additional dust beyond 
$\sim$1 kpc, highlighting the importance of magnetic-field structure and grain alignment in shaping the observed polarization.

\section{\textit{Planck} Data Processing and Polarization Analysis}\label{app:planck}

We used the 353 GHz \textit{Planck} Stokes maps ($Q$ and $U$), together with the corresponding 
variance maps ($QQ$ and $UU$), from the \textit{Planck} legacy release\footnote{The data were obtained 
from \url{https://pla.esac.esa.int/}. These products follow the IAU convention, in which the 
polarization angle is measured from the north Galactic pole and increases counterclockwise 
when looking at the source.} \citep{Planck_I:2020,Planck_III:2020}. These maps provide full-sky 
measurements of polarized thermal dust emission with well-characterized noise properties.

The native angular resolution of the \textit{Planck} data ($\mathrm{FWHM}=4.8'$) was degraded to a common 
resolution of $10'$ in order to match the spatial scale of the cells used to compute the mean stellar 
polarization (Appendix~\ref{app:ave}) and to improve the signal-to-noise ratio of the extended emission. 
This was achieved by convolving all Stokes and variance maps with the same Gaussian kernel.

The polarization position angle of the \textit{Planck} emission was derived from the Stokes parameters as
\begin{equation}
\psi = \frac{1}{2}\,\arctan2(U, Q).
\end{equation}

In this work, we assume that the optical starlight polarization angle, $\theta_\mathrm{stellar}$, traces 
the orientation of the plane-of-sky component of the magnetic-field, $\langle B_\mathrm{pos} \rangle$. 
Conversely, the polarization angles measured by \textit{Planck} are assumed to be perpendicular to the magnetic 
field direction. To enable a direct comparison between both tracers, the \textit{Planck} polarization angles were 
rotated by $90^\circ$ and transformed from Galactic to equatorial coordinates, yielding
\begin{equation}
\theta_{\rm Planck} = \psi + \frac{\pi}{2} - \Delta\theta(l,b),
\end{equation}
where $\Delta\theta(l,b)$ is the angle between the directions of Galactic north and equatorial north at 
the position of each line of sight. Using standard relations for spherical triangles (epoch J2000.0), this 
quantity can be written as \citep{Corradi:1998}
\begin{equation}
\Delta\theta(l,b) = \arctan2 \left [ 
\frac{\cos(l - 32\fdg9)}{\cos b \cot 62\fdg9 - \sin b \sin(l - 32\fdg9)} 
\right ],
\end{equation}
where $l$ and $b$ are the Galactic coordinates.

The uncertainty in the polarization angle was estimated by propagating the variances of the Stokes 
parameters, assuming that $Q$ and $U$ are uncorrelated \citep[e.g.,][]{Planck_XI:2020},
\begin{equation}
\sigma_{\psi} = 28\fdg65 \, \frac{\sqrt{Q^2\,\sigma_U^2 + U^2\,\sigma_Q^2}}{Q^2 + U^2},
\end{equation}
where $\sigma_Q^2$ and $\sigma_U^2$ correspond to the $QQ$ and $UU$ maps, respectively.

Finally, the \textit{Planck} polarization angles were compared with the stellar-polarization angles 
$\theta_\mathrm{stellar}$. The relative orientation was computed as
\begin{equation}
\Delta\theta = \frac{1}{2}\,\arctan2\left[\sin(2\theta_\mathrm{stellar} - 2\theta_{\rm Planck}), 
\cos(2\theta_\mathrm{stellar} - 2\theta_{\rm Planck})\right],
\end{equation}
which properly accounts for the $180^\circ$ degeneracy of polarization angles and restricts the 
resulting differences to the range $[-90^\circ, 90^\circ]$.

\section{Stellar-polarization Averaging Procedure}\label{app:ave}

We divided the observed region into a uniform grid of $10'\times10'$ cells. For each 
cell, we selected only stars satisfying $P/\sigma_{P} \ge 5$, and we required a minimum of 10 
stars per cell to ensure reliable statistics. The mean Stokes parameters were computed using 
inverse-variance weighting:

\begin{equation}
  \langle Q \rangle = \frac{\sum (q_i/\sigma_i^2)}{\sum \sigma_i^{-2}},  
\end{equation}

\begin{equation}
  \langle U \rangle = \frac{\sum (u_i/\sigma_i^2)}{\sum \sigma_i^{-2}},  
\end{equation}
where $q_i$ and $u_i$ are the normalized Stokes parameters of each star, and $\sigma_i$ is the 
corresponding uncertainty.

The mean polarization degree and its uncertainty follow from:

\begin{equation}
  \langle P \rangle = \sqrt{\langle Q \rangle^2 + \langle U \rangle^2},  
\end{equation}

\begin{equation}
   \delta P = \frac{\delta Q|\langle Q \rangle| + \delta U |\langle U \rangle|}{\langle P \rangle},
\end{equation}
where $\delta Q$ and $\delta U$ represent the standard deviations of the weighted means.

The mean polarization angle is computed as:
\begin{equation}
    \theta_{\langle P\rangle} = \onehalf 
    \arctan\!\left(\langle U\rangle/\langle Q\rangle \right).
    \label{eq:theta}
\end{equation}

\section{\HI\ Filament orientation.}\label{app:rht}

We applied the Rolling Hough Transform (RHT) analysis developed by \citet{Clark:2014}
to the \HI\ self-absorption map of the R--C cloud at 
$v = 4.95\,\mathrm{km\,s^{-1}}$ in order to highlight the most elongated and coherent 
linear features and to facilitate comparison with the magnetic-field orientation 
derived from stellar polarization.

To assess the robustness of the extracted filament orientations, we
explored a broad range of RHT parameter values around our adopted
choices for the smoothing kernel diameter ($D_K$), the rolling window
diameter ($D_W$), and the intensity threshold ($Z$), and our choices 
were $D_K$ = 75 pixels (43.7 arcmin), $D_W$ = 15 pixels (8.7 arcmin), 
and $Z = 80\%$ of the peak intensity of the map. Although variations
in these parameters affect the continuity and visibility of
lower-contrast structures, we find that the orientations of the
densest and most coherent filaments remain stable within a few
degrees. This demonstrates that the filament orientations used in our
comparison are not driven by fine-tuning of the RHT parameters, but
instead reflect intrinsic morphological properties of the
\HI\ structures.

We emphasize that the comparison with the stellar-polarization data is
dominated by these high-contrast, well-defined filaments. Lower-contrast
or more complex structures identified by the RHT can exhibit a broader
range of orientations and contribute to the overall angular dispersion,
but they play a secondary role in the filament--field alignment discussed
in this work. This behavior is consistent with previous applications of
the RHT to diffuse \HI, in which the most elongated filaments
preferentially trace the plane-of-sky magnetic field.

\section{Angular Dispersion Function analysis}\label{app:adf}

Estimating the plane-of-sky magnetic-field strength in interstellar clouds commonly relies on 
the Davis–Chandrasekhar–Fermi (DCF) approach \citep{Davis:1951, Chandrasekhar:1953}, which 
links angular perturbations of the magnetic field to the dynamical influence of turbulence. 
In its classical form, the method assumes that irregularities in the polarization field trace the 
response of magnetized gas to turbulent motions. The field strength is written as
\begin{equation}\label{eq:dcf}
    B_{\rm pos} \simeq Q\,\sqrt{4\pi\rho}\,
        \frac{\sigma_{v}}{\sigma_{\theta}},
\end{equation}
where $\rho$ is the gas density, $\sigma_{v}$ the nonthermal velocity dispersion,  and
$\sigma_{\theta}$ the turbulent dispersion of polarization angles. $Q$ is a correction 
factor of order unity  (typically $Q \sim 0.5$  \citep{Ostriker:2001, Crutcher:2004}) that 
accounts for  averaging effects and deviations from ideal assumptions.

Directly estimating $\sigma_{\theta}$ from the raw distribution of polarization angles, however, 
can severely bias the inferred field strength because large-scale, nonturbulent structure 
contributes to the observed dispersion. To isolate the turbulent component, we adopt the Angular 
Dispersion Function (ADF) technique \citep{Hildebrand:2009}, which quantifies the scale dependence 
of angle differences while explicitly separating ordered and turbulent contributions. The ADF is 
defined as
\begin{equation}\label{eq:phi_squared}
    \langle \Delta\Phi^{2}(\ell)\rangle 
    = \frac{1}{N(\ell)}\sum_{i=1}^{N(\ell)}
      \big[\,\Phi(x) - \Phi(x + \ell)\,\big]^{2},
\end{equation}
where $\Phi(x)$ is the polarization angle at position $x$ and $N(\ell)$ is the number of independent 
vector pairs separated by distance $\ell$.

Following the standard treatment, we model the magnetic field as the sum of two statistically 
independent components: a smoothly varying large-scale field $B_{0}(\mathbf{x})$ with characteristic 
scale $d$, and a turbulent component $B_{t}(\mathbf{x})$ characterized by a correlation 
length $\delta$. In the regime $\delta < \ell \ll d$, the observed ADF is well approximated by
\begin{equation}
    \langle \Delta\Phi^{2}(\ell)\rangle_{\rm tot}
    \simeq b^{2} + m^{2}\ell^{2} + \sigma^{2}_{\rm M}(\ell),
\end{equation}
where $b$ measures the turbulent angular dispersion, the term $m^{2}\ell^{2}$ captures the gradual 
rise introduced by the ordered field, and $\sigma^{2}_{\rm M}(\ell)$ accounts for measurement 
uncertainties in the polarization angles. The latter is computed from the propagated uncertainties of 
each pair and subtracted from the total ADF to recover the intrinsic function.

A fit to the distribution obtained by calculating the equation (\ref{eq:phi_squared}) in the appropriate 
spatial range produces the zero-separation intercept $b^{2}$, which provides the unbiased estimate of the 
turbulent angular dispersion required in the DCF formula. This procedure ensures that only the stochastic 
component of the magnetic field contributes to the inferred strength, avoiding contamination by the 
coherent morphology of the large-scale field.

The turbulent parameter $b$ is associated to the ratio between the turbulent and the large-scale magnetic 
fields as
\begin{equation}
    \frac{\langle B_t^2 \rangle^{1/2}}{B_0} = \frac{b}{\sqrt{2 - b^2}}.
\end{equation}
Assuming that the dispersion in polarization angles $\sigma_\theta$ is given by
\begin{equation}
    \sigma_\theta = \frac{\langle B_t^2 \rangle^{1/2}}{Bo}, 
\end{equation}
Equation\,(\ref{eq:dcf}) became
\begin{eqnarray}
     B_0 & \simeq &Q\sqrt{(2 - b^2) 4 \pi \rho} \frac{\sigma_v}{b} \\
         & \simeq & Q\sqrt{8\pi\rho} \frac{\sigma_v}{b} 
\end{eqnarray}
where the last equation applies when $b^2 \ll 2$.

We adopted the same dispersion velocity $\sigma_v = 1.4$ km s$^{-1}$ used in previous
estimates of the magnetic-field strength acting on the R--C cloud; however, instead
of adopting the same volume mass density for the whole mapped cloud, we estimated the following.
The interstellar absorption, $A_\mathrm{V}$, of the stars used to estimate the dispersion of the 
polarization angles was retrieved from the \texttt{StarHorse} catalog \citep{Anders:2022} and the average value was
converted into the column density of \HI\ through the standard gas-to-extinction relation 
$N_\mathrm{H} \simeq 1.87 \times 10^{21} A_\mathrm{V}$ cm$^{-2}$ mag$^{-1}$ \citep{Bohlin:1978}.
To avoid contamination from background clouds to the estimated value of mean interstellar
absorption, we limited our sample of stars to a distance of 1 kpc. Beyond that distance, the 
effects of the dust material associated with the Sagittarius-Carina arm are noticeable on the
distribution of interstellar absorption.

To obtain an estimate of the volume mass density, we adopted the upper limit of 3.5 pc for
 the thickness of the R--C cloud suggested by  \citet{Roshi:2011}. Although this value lies 
 within the 1–5 pc range proposed by \citet{Crutcher:1984}, it remains the parameter that contributes 
 the largest uncertainty to the derived magnetic-field strength. To account for this, we adopted 
 an {\it ad hoc} uncertainty of $\pm$0.5 pc on the cloud thickness. It is also important to emphasize that 
 the suggested thickness by \citet{Roshi:2011} is based on radio recombination line (RRL) modeling, which traces 
 partially ionized gas, whereas the extinction and polarization measurements primarily probe the neutral, 
 dust-bearing medium. As a result, the assumed thickness may not strictly correspond to the same 
 physical component responsible for the observed extinction and polarization and should therefore 
 be regarded as an approximate, order-of-magnitude estimate.

The uncertainties in the derived physical quantities were estimated using a Monte Carlo propagation 
of errors. The input parameters $A_V$, $b$, and $l$ were assumed to follow Gaussian distributions centered 
on their measured values with standard deviations equal to their observational uncertainties. A large number 
of random realizations were generated for each parameter. Nonphysical values (negative $A_V$, $b$ and $l$) 
were discarded before proceeding with the calculations. For each remaining realization, 
the code computes the intermediate quantities $N_H$, $n_H$,  and $\rho$, the turbulent-to-ordered field ratio 
$\langle B_t ^2 \rangle^{1/2}/B_0$, and the plane-of-sky magnetic field strength $B_\mathrm{pos}$.
The resulting distributions of these derived quantities reflect the propagation of the input uncertainties through 
the nonlinear equations. The final reported values correspond to the mean of each simulated distribution, while the 
uncertainties are taken as the standard deviation of those distributions, providing an estimate of the propagated 
error in each physical parameter.

Once the plane-of-sky magnetic-field strength is estimated from the ADF
analysis, the corresponding Alfv\'en speed can be computed as
\begin{equation}
v_A = \frac{B}{\sqrt{4\pi\rho}},
\end{equation}
where $B$ is the magnetic-field strength and $\rho$ is the volume mass
density of the gas. For completeness, we note that for a population of randomly oriented 
magnetic fields the expectation value of the plane-of-sky component is
$\langle B_\mathrm{pos} \rangle = (\pi/4)\,|\mathbf{B}|$ \citep{Crutcher:2004}. Given the
highly ordered magnetic-field geometry observed in the R--C cloud,
we do not attempt to infer the total magnetic-field strength from $B_\mathrm{pos}$.

In Figs.\,\ref{fig:adf_plots}  we present the measured ADF 
($\langle \Delta\Phi^2(\ell)\rangle - \sigma_\mathrm{M}^2$), as a function of the angular separation ($\ell$)
together with the best-fit model for each of the regions defined in Fig.\,\ref{fig:map_fil}. 
The three regions display distinct ADF behaviors, indicating variations in the relative contributions 
of turbulent and ordered magnetic-field components. Field B shows a clear increase of the ADF with 
separation over a wide range of scales, reflecting the presence of a coherent large-scale magnetic structure, 
but it also exhibits the largest turbulent-to-ordered magnetic-field ratio. In contrast, Fields A and C show smaller values 
$\langle B_t^2 \rangle^{1/2}/B_0$, indicating comparatively weaker turbulent fluctuations with respect to the 
ordered field. Field A presents an intermediate ADF behavior, while Field C shows a flatter distribution at larger 
separations.

Overall, these results suggest that although a large-scale magnetic field is present in all three 
regions, the relative importance of turbulence is highest in Field B, whereas Fields A and C are 
characterized by similar and lower levels of turbulent perturbations relative to the ordered magnetic field. 
This demonstrates that angular dispersion analysis provides a robust diagnostic of the interplay 
between turbulent fluctuations and the large-scale magnetic-field structure across different regions.

An important implication of our results concerns previous estimates of the magnetic-field strength in the 
R--C cloud based on the RHT analysis of \HI\ filamentary structures. While the RHT has proven to be a 
powerful tool for identifying linear features in atomic gas and for revealing the connection between 
filamentary morphology and magnetic-fields, its interpretation relies on the assumption that filament 
orientations accurately trace the underlying plane-of-sky magnetic-field. Within this framework, the 
angular dispersion of filament orientations is commonly used as a proxy for the magnetic-field dispersion.
As illustrated in Fig.\,\ref{fig:map_fil} (see also Figs.\,\ref{fig:fields_detail}), 
the dispersion inferred from the full population of RHT-identified filaments is systematically larger 
than that measured directly from stellar polarization. The densest and most coherent filaments 
are well aligned with the local magnetic-field; however, the RHT also identifies lower-contrast 
and morphologically complex structures whose orientations may not reliably trace the underlying 
field geometry. When all detected structures are treated on equal footing, these features broaden 
the inferred angular distribution and consequently bias the magnetic-field strength toward lower 
values, given the inverse relationship between field strength and angular dispersion.
This behavior should not be interpreted as a limitation of the RHT itself, but rather as a 
natural consequence of the complex, multiphase structure of the \HI\ medium, where not all 
identified structures are expected to trace the same magnetic-field component. In this context, 
stellar-polarimetry provides a complementary and more direct probe of the magnetic-field orientation 
through grain alignment along discrete lines of sight, enabling a more reliable determination of the 
intrinsic angular dispersion in magnetically dominated regions.


\end{document}